\documentclass[
  aps,
  prl,
  preprint,
  superscriptaddress,
  nobibnotes
]{revtex4-2}

\usepackage{graphicx}
\usepackage{amsmath,amssymb,amsfonts}
\usepackage{mathrsfs}
\usepackage{xcolor}
\usepackage{textcomp}
\usepackage{booktabs}
\usepackage{multirow}

\setcitestyle{super,sort&compress}


\def\bibinfo#1#2{%
  \begingroup
  \def\bibtype{#1}%
  \def\journaltype{journal}%
  \ifx\bibtype\journaltype
    \textit{#2}%
  \else
    #2%
  \fi
  \endgroup
}

\usepackage[hidelinks]{hyperref}
\usepackage[nameinlink]{cleveref}
\crefname{figure}{Fig.}{Figs.}

\newlength{\defaultparindent}
\AtBeginDocument{%
  \setlength{\defaultparindent}{\parindent}%
  \setlength{\parindent}{2\defaultparindent}%
}

\makeatletter
\renewcommand\section{%
  \@startsection
    {section}{1}{\z@}%
    {0.8cm \@plus1ex \@minus .2ex}%
    {0.35cm}%
    {\normalfont\small\bfseries\centering}%
}
\renewcommand\subsection{%
  \@startsection
    {subsection}{2}{\z@}%
    {0.8cm \@plus1ex \@minus .2ex}%
    {0.35cm}%
    {\normalfont\small\bfseries\centering}%
}
\renewcommand\subsubsection{%
  \@startsection
    {subsubsection}{3}{\z@}%
    {0.8cm \@plus1ex \@minus .2ex}%
    {0.35cm}%
    {\normalfont\small\itshape\centering}%
}
\makeatother

\makeatletter
\renewcommand\frontmatter@abstractheading{}
\renewcommand\frontmatter@abstractfont{\bfseries}

\long\def\@makecaption#1#2{%
  \par
  \vskip\abovecaptionskip
  \begingroup
   \small\rmfamily
   \dimen@=\f@size\p@
   \baselineskip=1.38\dimen@  
   \sbox\@tempboxa{%
    \let\\\heading@cr
    \@make@capt@title{#1}{#2}%
   }%
   \@ifdim{\wd\@tempboxa >\hsize}{%
    \begingroup
     \samepage
     \flushing
     \let\footnote\@footnotemark@gobble
     \@make@capt@title{#1}{#2}\par
    \endgroup
   }{%
     \global \@minipagefalse
     \hb@xt@\hsize{\hfil\unhbox\@tempboxa\hfil}%
   }%
  \endgroup
  \vskip\belowcaptionskip
}%

\renewcommand\refname{References}
\renewcommand\bibsection{\section*{\refname}\@nobreaktrue}
\makeatother

\begin{document}
\pagestyle{plain}

\title{Topological terahertz fields from tunable vectorial currents}

\author{Jacob Pettine}
\thanks{These authors contributed equally to this work.}
\affiliation{Center for Integrated Nanotechnologies, Los Alamos National Laboratory, Los Alamos, New Mexico 87545, USA}
\affiliation{Department of Physics, Massachusetts Institute of Technology, Cambridge, Massachusetts 02139, USA}

\author{Tae Gwan Park}
\thanks{These authors contributed equally to this work.}
\affiliation{Center for Integrated Nanotechnologies, Los Alamos National Laboratory, Los Alamos, New Mexico 87545, USA}
\affiliation{Department of Physics and Photon Science, Gwangju Institute of Science and Technology, Gwangju 61005, Republic of Korea}

\author{Yanhe Lyu}
\affiliation{Centre for Disruptive Photonic Technologies, School of Physical and Mathematical Sciences, Nanyang Technological University, Singapore 637371, Singapore}

\author{Wangke Yu}
\affiliation{Centre for Disruptive Photonic Technologies, School of Physical and Mathematical Sciences, Nanyang Technological University, Singapore 637371, Singapore}

\author{John B. Maier}
\affiliation{Department of Physics, Massachusetts Institute of Technology, Cambridge, Massachusetts 02139, USA}

\author{Matthew Yeung}
\affiliation{Department of Physics, Massachusetts Institute of Technology, Cambridge, Massachusetts 02139, USA}

\author{Rujuta Sane}
\affiliation{Department of Physics, Massachusetts Institute of Technology, Cambridge, Massachusetts 02139, USA}

\author{Chun-Chieh Chang}
\affiliation{Center for Integrated Nanotechnologies, Los Alamos National Laboratory, Los Alamos, New Mexico 87545, USA}

\author{Nuh Gedik}
\affiliation{Department of Physics, Massachusetts Institute of Technology, Cambridge, Massachusetts 02139, USA}

\author{Yijie Shen}\email{yijie.shen@ntu.edu.sg}
\affiliation{Centre for Disruptive Photonic Technologies, School of Physical and Mathematical Sciences, Nanyang Technological University, Singapore 637371, Singapore}
\affiliation{School of Electrical and Electronics Engineering, Nanyang Technological University, Singapore 639798, Singapore}

\author{Hou-Tong Chen}\email{chenht@lanl.gov}
\affiliation{Center for Integrated Nanotechnologies, Los Alamos National Laboratory, Los Alamos, New Mexico 87545, USA}

\begin{abstract}
\setlength{\parindent}{2em}
Future information processing calls for photonic architectures with greater capacity, intrinsic resilience to noise, and direct integration with quantum materials and devices. Topological light pulses have recently emerged as promising candidates to meet these demands, yet the complexity underlying their unique properties also makes them challenging to generate, tune, and fully resolve in space and time. Here, we generate and actively tune topological light fields through transient vectorial currents in an optoelectronic metasurface. Incident linear polarization steers the local direction of photocurrents in graphene via nanoplasmonic resonators with triangular symmetry, while a global winding pattern enables continuous tuning between compressional and circulating charge flows. Under ultrafast excitation, these currents radiate single-cycle terahertz toroidal pulses spanning electric- and magnetic-dominant modes with dynamically evolving skyrmion textures. Full spatiotemporal mapping and reconstruction provide a complete visualization of the toroidal structure of light, directly tracing the closed field surfaces while continuously tracking skyrmion number and helicity throughout a single cycle. These results connect two-dimensional vectorial currents to rich three-dimensional electromagnetic field structures, while establishing a versatile integrated platform for higher-dimensional photonic information processing and emerging material spectroscopies.
\end{abstract}

\maketitle
\thispagestyle{plain}

\section{Introduction}\label{sec1}

Structured light with spatially varying polarization, amplitude, and phase expands the photonic degrees of freedom available for classical and quantum information encoding\cite{Erhard2018, shen2019optical, Forbes2025, he2022towards}, while also opening new routes for probing symmetry and emergent order in condensed matter systems\cite{Rosen2022, Rouxel2022, cheng2025metrology}. These applications are closely connected, as material symmetries, topology, and dynamics can be embedded in structured light-matter interactions, which provide unique channels for both signal transduction and spectroscopic readout. Within the terahertz (THz) spectral range, effects such as helical dichroism of vortex modes\cite{Rouxel2022, Forbes2021} arise on the natural energy scale of correlations and collective excitations in quantum matter\cite{Sirenko2019, Sirenko2021}. Additional topological degrees of freedom can be introduced in vectorial light, supporting exotic field polarization textures like skyrmions\cite{Shen2024, shen2025free}. In the few-cycle limit, broadband THz vector fields can further exhibit toroidal topology, characterized by closed field loops, energy backflow, and intrinsically nonseparable space-time dynamics\cite{Hellwarth1996, Shen2022, Zdagkas2022}. The topological structure is resilient against perturbations\cite{Wang2024LSA, Guo2026NatCommun, Wang2024APR} and supports higher-dimensional information encoding based on topological invariants\cite{Shi2026, zhang2026skyrmions, Chen2026NatElectron}, with recent extensions into the quantum regime\cite{Forbes2025, Kam2026eLight}. The same rich field structure also opens new vectorial and multipolar coupling channels to matter, including out-of-plane and magnetic-dipole excitations\cite{Li2012, Kasperczyk2015}, toroidal moments\cite{Papasimakis2016}, and nonradiating anapole current configurations\cite{Savinov2019}.

Various methods exist for generating THz pulses with tunable vector polarization\cite{Petrov2022, Imai2012, Deveikis2022}, also providing direct routes to THz vortices\cite{Petrov2022, Imai2014}. Among these, exchange-biased spintronics\cite{Wang2024}, coherent control\cite{Jana2024}, and nonlinear metasurfaces\cite{Zdagkas2022, Niu2026} have emerged as versatile platforms for spatially patterning ultrafast bound and free charge currents and shaping the resulting THz emission. However, active control over complex THz vector fields from a single compact source remains limited, particularly where such tuning can be achieved through a simple optical input parameter. Unambiguous spatiotemporal characterization also demands precisely engineered source currents and sensitive field mapping, leaving the three-dimensional (3D) topology and ultrafast evolution of toroidal light only partially resolved.

Here, we introduce plasmonic gold metasurfaces with $C_3$ nanoscale symmetry to drive vectorial currents in underlying graphene, with local directionality determined by the linear polarization of the incident optical field. Global winding structure of the metasurface converts this local control into continuous tunability between radial compressional flows with charge buildup and azimuthal circulating flows with current vorticity. Under ultrafast excitation, these transient currents emit THz vector pulses, generating toroidal topologies within the 3D electric and magnetic fields. Sensitive spatiotemporal mapping and 3D field reconstruction directly reveal the closed toroidal structures and their sub-picosecond evolution through all four combinations of skyrmion number and helicity. Tuning the current texture continuously transforms the emission between transverse magnetic (TM) and transverse electric (TE) toroidal modes and their sign-reversed counterparts. We thus demonstrate an efficient, compact, and material-integrated source of toroidal light pulses with a simple optical tuning knob, in which 3D free-space vector topology is inherited from ultrafast two-dimensional (2D) charge currents.

\section{Continuously tunable vectorial nanocurrents}\label{sec2}

\begin{figure*}[b]
    \centering
    \includegraphics[width=0.95\textwidth]{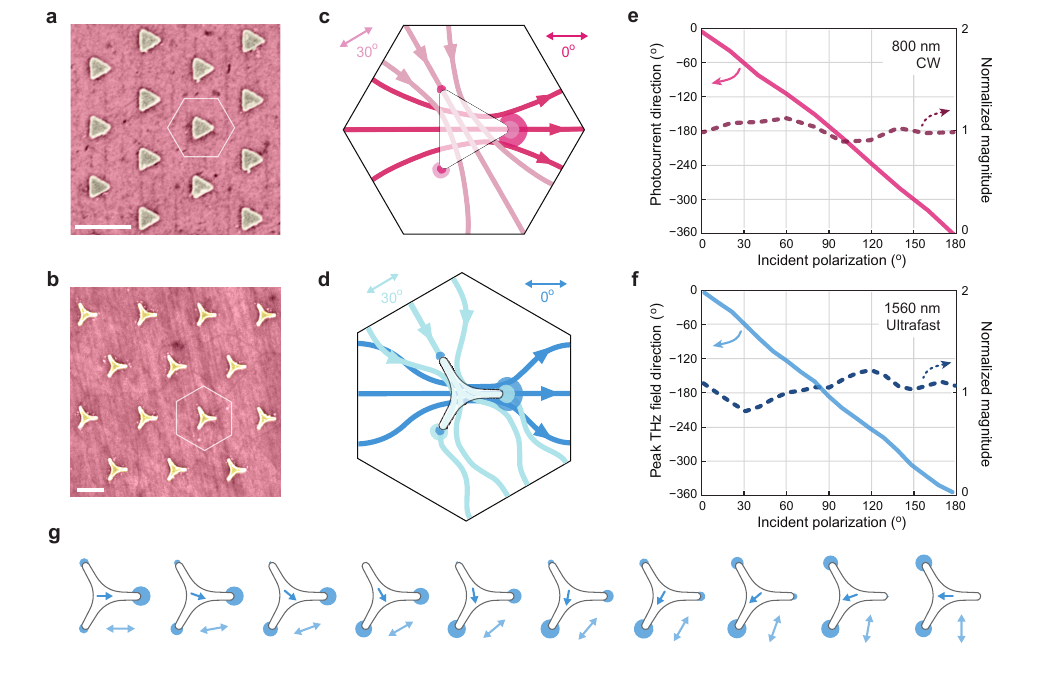}
    \caption{\textbf{Full directional control over nanoscale currents in $C_3$ metasurfaces.}
    Scanning electron micrographs of 
    \textbf{a}, triangular (resonance at 800\,nm) and 
    \textbf{b}, three-petal (resonance at 1560\,nm) gold nanoantenna arrays on graphene. Scale bars, 500 nm. \textbf{c},\textbf{d}, Charge flow lines within the unit cells for $0^\circ$ (darker lines) and $30^\circ$ (lighter lines) incident polarization, simulated within a hydrodynamic approximation. Solid fills around the nanoantenna apexes highlight simulated regions with a field enhancement of 10-fold or greater. 
    \textbf{e}, Measured photocurrent in the triangular metasurface under resonant continuous-wave excitation at 800\,nm. 
    \textbf{f}, Peak THz field emitted from the three-petal metasurface under resonant excitation with 1560\,nm femtosecond laser pulses. In \textbf{e} and \textbf{f}, dashed and solid lines represent magnitude and direction, respectively, as a function of the incident polarization angle.
    \textbf{g}, Evolution of local current direction (center arrows) and field enhancements for a 0$^\circ$--90$^\circ$ incident polarization series (double arrows).}
    \label{fig1}
\end{figure*}

Currents emerge as a general consequence of spatial symmetry breaking, with mechanisms such as internal photoemission at metal-semiconductor junctions\cite{Taghinejad2023} or internal boundary scattering and ratchet effects in etched magnetic films\cite{Matsubara2022} yielding net directionality. Gold resonators induce particularly strong responses in graphene under both continuous-wave\cite{Wei2020} and ultrafast\cite{Pettine2024} excitations due to resonant enhancement of junction photothermoelectric effects\cite{Shautsova2018NatCommun} at nanoscale plasmonic hot spots. Hot carriers excited in graphene diffuse away from these hot spot regions with a directionality imparted by the spatial variation in the chemical potential and corresponding Seebeck coefficient across the gold--graphene interface. As shown in \cref{fig1}, the unequal field enhancement at the three hot spots around each nanoantenna results in a net directional photocurrent within the unit cell, which can be driven in any direction with the same magnitude by rotating the linear polarization of the incident light. This behavior persists across different resonator geometries (triangle vs. three-petal; \cref{fig1}a,b), resonant excitation wavelengths (800\,nm vs. 1560\,nm), relative nanostructure orientations within the unit cell ($0^\circ$ vs. $30^\circ$), excitation timescales (continuous-wave vs. femtosecond pulses), and spatial scales (local diffraction-limited focused laser spot vs. global excitation of the full metasurface). Under the same average power, both continuous-wave and pulsed excitations yield similar average currents read out directly via electrodes, while pulsed excitation also generates strong THz emission due to the sub-picosecond rise and decay dynamics of the charge flow\cite{Pettine2024}. 

Spatial symmetry serves as the unifying design principle determining the $j_i = \sigma_{ijk}^{(2)} E_j(\omega)E_k^*(\omega)$ photocurrent response across different material platforms\cite{Pettine2023}, with the magnitude of $\sigma_{ijk}^{(2)}$ depending upon the specific microscopic mechanism. The ability to continuously tune photocurrent directionality from $0^\circ$ to $360^\circ$ via incident optical polarization is a manifestation of Neumann's principle\cite{Powell2010}, whereby any system with $C_3$ symmetry will exhibit an isotropic second-order response about the corresponding symmetry axis. Calculated field enhancements and resulting flow lines modeled in the hydrodynamic regime (Methods) are shown in \cref{fig1}c,d for different incident polarizations. The measured local photocurrent response in \cref{fig1}e and global ultrafast THz emission in \cref{fig1}f verify the continuous directional control---further illustrated in \cref{fig1}g---with only small variations in magnitude attributed to fabrication imperfections, graphene nonuniformity, and other experimental errors. Although both readouts are effectively global, the local nature of the vectorial control around each individual nanoantenna is clarified by the diffraction-limited excitation for the photocurrent measurement, where only a few nanoantennas fall within the laser spot.

The strength of the ultrafast photocurrent response is indicated simultaneously by the dc readout and the THz emission, with the latter often even exceeding the emission from a conventional 1-mm-thick $\langle110\rangle$ ZnTe THz source under the same optical fluence\cite{Pettine2024}. Nonlinear rectification is also built into the internal response of asymmetric gold nanostructures, as utilized recently for THz generation\cite{Luo2014NatCommun, Zdagkas2022, McDonnell2022}. However, when compared against ZnTe standards, the THz fields measured here are found to be one to two orders of magnitude stronger than those produced by the internal rectification mechanism in nonlinear metasurfaces. 

\begin{figure*}[b]
    \centering
    \includegraphics[width=0.95\textwidth]{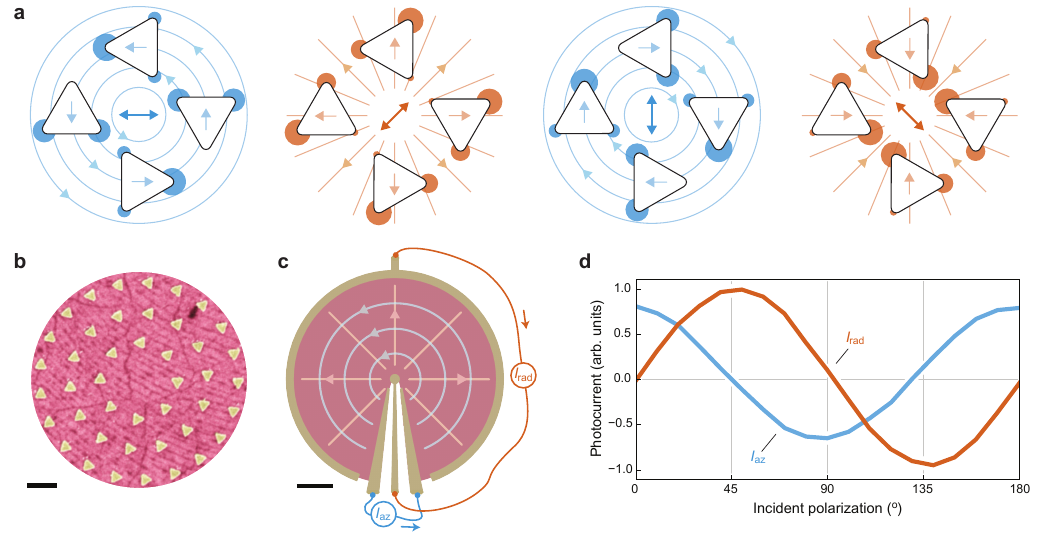}
    \caption{\textbf{Continuously tunable vectorial currents in a winding metasurface.}
    \textbf{a}, Illustration of the evolution between circulating azimuthal modes (blue) and compressional radial modes (orange) for a winding metasurface. 
    \textbf{b}, Scanning electron micrograph of the central region of a 1-mm-diameter metasurface resonant at 800\,nm, before patterning electrodes. Scale bar, 500 nm. 
    \textbf{c}, Optical micrograph of Corbino device after patterning of radial and azimuthal electrodes, with regions where the graphene is etched away shown as solid white. Scale bar, 200 µm. 
    \textbf{d}, Measured radial (orange) and azimuthal (blue) photocurrents as a function of the linear polarization angle of the incident 800\,nm femtosecond laser.}
    \label{fig2}
\end{figure*}

The versatility conferred by the combined light-matter response allows simple, linearly-polarized light fields to drive more complex vectorial currents with a high degree of tunability when the metasurface itself exhibits nonuniform spatial patterning. As illustrated in \cref{fig2}a, winding triangles by $120^\circ$ around the center of the metasurface (\cref{fig2}b) leads to a tunable counter-clockwise azimuthal ($0^\circ$ incident polarization), outward radial ($45^\circ$), clockwise azimuthal ($90^\circ$), and inward radial ($135^\circ$) photocurrent progression, with all intermediate conditions continuously sampled. We read these global vectorial currents out directly with radial and azimuthal electrodes (\cref{fig2}c) and find nearly perfect sinusoidal responses for the two channels, offset by the expected $45^\circ$ in incident polarization (\cref{fig2}d). This indicates the high quality of the patterned current response, which facilitates the continuously tunable vectorial structure of the THz pulses emitted under ultrafast excitation, described below. Small dc current offsets may be attributed to relatively minor photothermoelectric current contributions from the gold electrodes.

We note that these currents also produce strong and tunable near-field distributions. For both radial and azimuthal modes, dc currents around 10\,nA are measured in \cref{fig2}d under modest 0.1\,µJ/cm$^2$ incident laser fluence. From the device geometry, the 80\,MHz laser repetition rate, and a $\sim$250\,fs current pulse duration, we estimate peak transient magnetic fields $B_z \approx$ 1--10 µT near the center of the circulating current loops (SI Section 3). Moreover, these metasurfaces can be driven with 1000-fold higher fluence before reaching the damage threshold of the gold antennas or the peak current density of the graphene\cite{Pettine2024}. Thus, neglecting the weakly sublinear scaling at these higher fluences, the millitesla range is expected to be accessible (Fig.\,S4). For compressional radial modes, charge buildup within the central metasurface region (before electrode patterning) leads to $E_z > 1~\mathrm{kV/cm}$ within a few microns of the surface (Fig.\,S6). We expect far larger fields will become possible at microscopic scales with further engineering of the photoresponse in higher-mobility exfoliated graphene\cite{Wei2020}, enhanced Seebeck coefficient differentials near charge neutrality\cite{Shautsova2018NatCommun, Pettine2024}, multilayer stacks for higher peak current density, and different hybrid material platforms. These strong, textured, and dynamically tunable near fields can be used to drive nearby material excitations, particularly in layered heterostructures. Furthermore, as we show next, the skyrmionic field textures encoded within the near-field $(E_x, E_y, E_z)$ and $(B_x, B_y, B_z)$ and the underlying photocurrent $(j_x, j_y,\nabla\cdot \textbf{j})$ and $(j_x,  j_y,\nabla\times\textbf{j})$ (SI Section 3), respectively, are inherited by the far-field nonparaxial vectorial profiles of the radiated THz fields.

\section{Tunable toroidal pulses}\label{sec3}

Toroidal light pulses are an emerging class of spatiotemporally textured waves that carry 3D electromagnetic topological structure at the single-cycle limit. Recent work demonstrated these modes for the first time using metasurfaces~\cite{Zdagkas2022}, with further characterizations across optical~\cite{vignjevic2026state}, THz~\cite{Jana2024, Niu2026}, and microwave domains~\cite{Wang2024APR, Wang2025SciAdv, Shi2026}. We now explicitly visualize and actively tune the ultrafast 3D electric and magnetic structure of toroidal light in the THz range. 

\begin{figure*}[b]
    \centering
    \includegraphics[width=\textwidth]{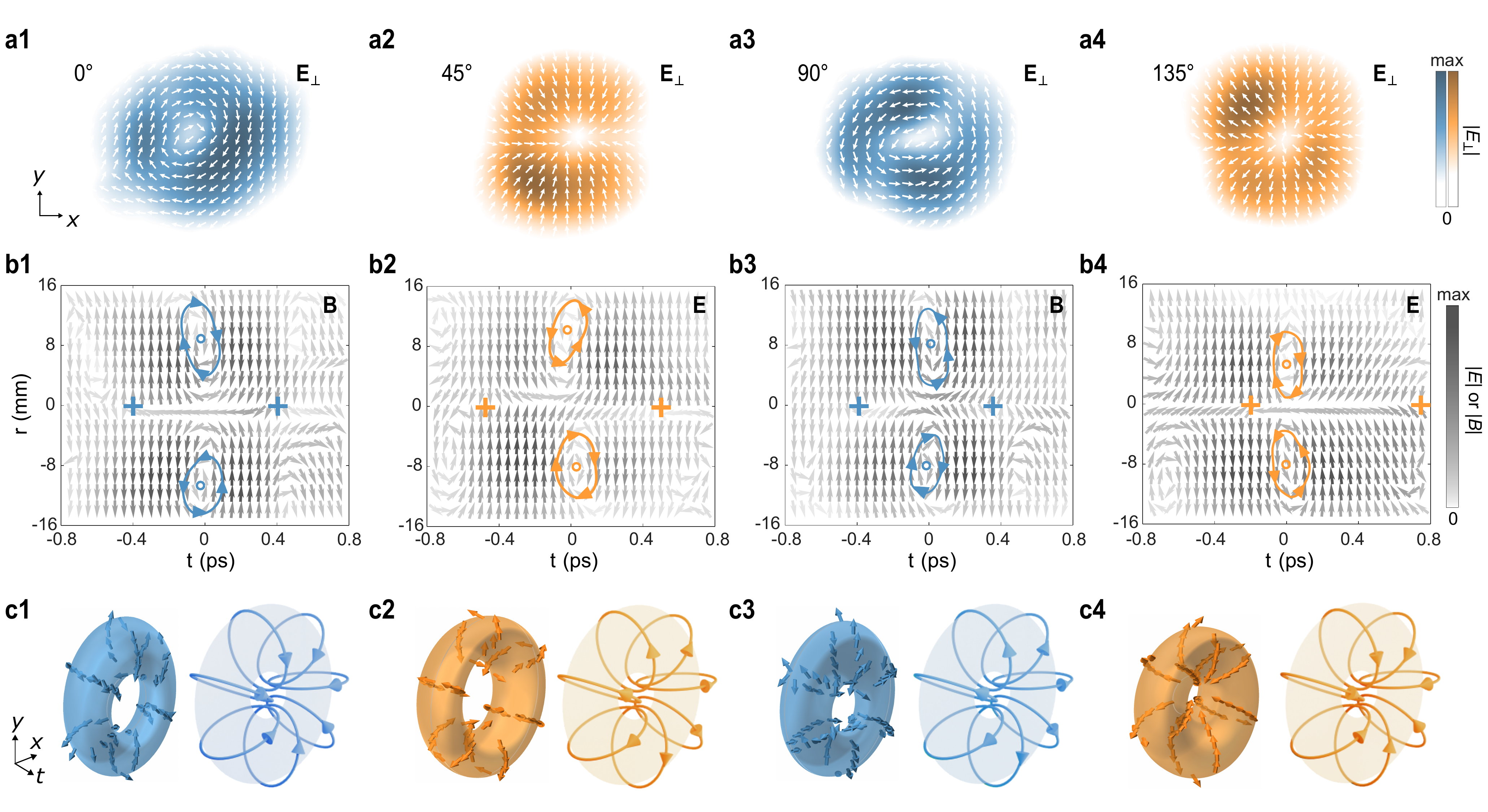}
    \caption{\textbf{Vortex tracing and three-dimensional reconstruction of tunable toroidal pulses.}
    \textbf{a1--a4}, Transverse electric field profiles for the leading THz pulse cycle (negative time delay) for incident linear polarization angles of 0$^\circ$, 45$^\circ$, 90$^\circ$, and 135$^\circ$, respectively. 
    \textbf{b1--b4}, Full spatiotemporal electromagnetic field evolution reconstructed from the experimentally measured transverse electric field traces. The four excitation states correspond to the magnetic field of TE, the electric field of TM, the magnetic field of negative TE, and the electric field of negative TM toroidal pulses. Grayscale color maps show the field magnitude, $|E|$ or $|B|$, with arrows indicating the field vectors in the $(r,t)$ plane, hollow circles and crosses mark the primary off-axis vortical field-null points and on-axis saddle points, respectively, while colored loops with arrows indicate the associated field circulation. 
    \textbf{c1--c4}, Full 3D magnetic (blue) and electric (orange) toroidal modes constructed from the vector-field loops for the four excitation states, shown alongside illustrations of the ideal toroidal modes.}
    \label{fig3}
\end{figure*}

A three-petal implementation of the winding metasurface is resonantly excited with a 1560\,nm femtosecond laser to generate THz vector pulses. Employing a sensitive scanning fiber-coupled photoconductive antenna, we map the THz emission as a function of time and lateral position for hyperspectral images of the THz field profiles (Methods), measured away from the focal plane for improved resolution (Fig.\,S2). Maps of the transverse electric field at the peak of the leading THz pulse cycle are shown in \cref{fig3}a, with additional cross-sections of the evolution through different input polarizations shown in Figs.\,S8--S11. Through these maps, we find that the polarization across a phase front follows the vectorial distribution of the corresponding photocurrent, with the central singularity characteristic of a doughnut beam. As the incident linear polarization is rotated between $0^\circ$, $45^\circ$, $90^\circ$, and $135^\circ$, the emitted THz vector field continuously evolves between clockwise azimuthal, inward radial, counter-clockwise azimuthal, and outward radial electric fields, respectively. This is consistent with the measured photocurrent evolution, considering that $\mathrm{\textbf{E}_{THz}} \propto -d\textbf{j}/dt$. The single-cycle nature of these pulses is evident in the time-domain traces, with Fourier transforms yielding a broad spectral response beyond 3\,THz (Fig.\,S3). 

From direct measurements of the two transverse electric-field components, $E_x(x,y,t)$ and $E_y(x,y,t)$, we retrieve the remaining fields by imposing the source-free Maxwell constraints on the measured angular spectrum. Briefly, the measured transverse fields are Fourier transformed in $(x,y,t)$, and for each propagating component with wavevector $\mathbf{k}=(k_x,k_y,k_z)$ we obtain the longitudinal electric field from Gauss's law, $\mathbf{k}\cdot\mathbf{E}=0$, giving $\widetilde{E}_z=-(k_x\widetilde{E}_x+k_y\widetilde{E}_y)/k_z$. The magnetic field is reconstructed from Faraday's law, $\widetilde{\mathbf{B}}=(1/\omega)\mathbf{k}\times\widetilde{\mathbf{E}}$, yielding $B_x$, $B_y$, and $B_z$. This procedure converts the measured transverse vector images into the full three-component spatiotemporal electromagnetic field, allowing the longitudinal $E_z$ and $B_z$ components and hence the TE/TM toroidal character to be fully resolved. To offset the large curvature of the converging wavefronts (Fig.\,S3), a flat-field phase correction is applied to the final three-component fields. 

The reconstructed electromagnetic fields reveal the toroidal topology inherited from the two orthogonal classes of vectorial photocurrents. \cref{fig3}b shows the trajectories of the primary spatiotemporal vector field in the $(r,t)$ plane for four pump polarization angles. The azimuthal current states at $0^\circ$ and $90^\circ$ generate magnetic-dominant TE toroidal pulses, shown through $\mathbf{B}(r,t)$, while the radial-current states at $45^\circ$ and $135^\circ$ generate electric-dominant TM toroidal pulses, shown through $\mathbf{E}(r,t)$. Rotating the pump linear polarization angle by $90^\circ$ reverses the current direction and thus inverts the toroidal field orientation, producing the negative-TE and negative-TM partners. Hollow circles mark the two primary off-axis vortical field-null points within the minor cross-sections of the torus, whereas crosses denote on-axis saddle points in the measured spatiotemporal vector field. The arrowed loops follow the dominant circulating field trajectories that define the toroidal skeleton. Retrieving these loops in three dimensions yields the closed toroidal structures shown in \cref{fig3}c, which closely resemble the ideal toroidal pulse structures. These measurements therefore establish that the winding metasurface does not merely emit transverse vector radiation, but indeed produces single-cycle toroidal THz pulses with continuous tuning between magnetic and electric toroidal topology. Though not shown here, intermediate states emitted by spiral currents exhibit additional helical topology with electric and magnetic field vectors winding around the torus\cite{Shi2026}.

To further characterize the toroidal character of these pulses, we numerically propagate the full electric field to the focal region using the spectral phase factor within the Fourier decomposition. Following this propagation, the retrieved pulse at the focal plane exhibits a nearly ideal single-cycle toroidal pulse structure, in close agreement with the analytic result (Fig.\,S15). A progressive redshift at larger radial position and associated isodiffraction behavior (Fig.\,S16) verify the space-time non-separability\cite{Zdagkas2022} and stable toroidal topology (SI Section 6).

\section{Ultrafast skyrmionic topological transitions}\label{sec4}

It has been theoretically predicted that toroidal light pulses evolve through four types of electromagnetic skyrmions, i.e., combinations of opposite polarity and opposite helicity\cite{Shen2021, Shen2024NDSTP, Shen2026STSO}. However, in previous experimental works~\cite{Niu2026, Jana2024, Wang2024APR, Wang2025SciAdv}, the dynamic transition of these properties has not yet been fully resolved or controlled. Here, we directly observe the ultrafast skyrmion dynamics within a single cycle by experimentally tracking the evolution of their topological charges and field textures. The generated toroidal pulses exhibit transitions among distinct skyrmion configurations within a single cycle, extending the control of skyrmions from static topological states to ultrafast spatiotemporal topological transitions.

\begin{figure*}[b]
    \centering
    \includegraphics[width=\textwidth]{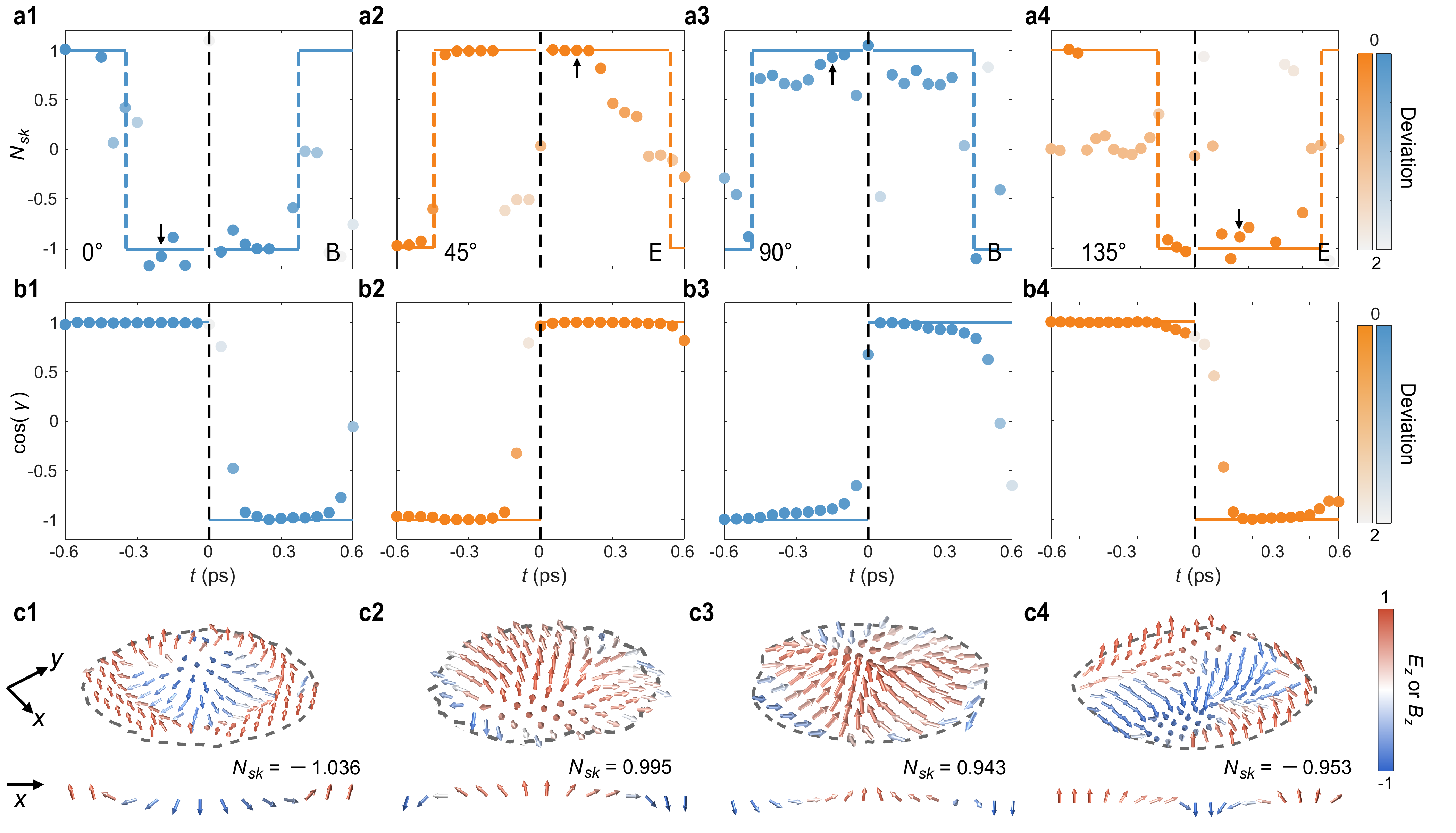}
    \caption{\textbf{Observation of ultrafast skyrmion evolution.}
    \textbf{a1--a4}, Skyrmion number $N_{\mathrm{sk}}$ as a function of time for the four measured toroidal states. Color maps clarify deviations from the ideal skyrmion number trajectories, shown as solid lines. The horizontal axes show the shifted time coordinate $t$, defined by setting the temporal position of the primary off-axis vortical field-null points as the zero-delay reference. 
    \textbf{b1--b4}, Measured helicity, represented by $\cos(\gamma)$, as a function of time.
    \textbf{c1--c4}, Measured skyrmion textures selected around the primary vector-field vortex of the four toroidal states, at time delays indicated by the arrows in \textbf{a1--a4}. Dashed loops outline the skyrmion boundary (Methods). $E_z$ and $B_z$ are scaled by normalization factors relative to their respective transverse field components for clarity.}
    \label{fig4}
\end{figure*}

We evaluate the spatiotemporal topologies of the generated toroidal excitations through the time-varying skyrmion number, $N_{\mathrm{sk}}$, and helicity, $\gamma$ (SI Section 4). $N_{\mathrm{sk}}$ quantifies the topological wrapping of the normalized field vector over the unit sphere, whereas $\gamma$ characterizes the angular offset between the in-plane field orientation and the radial direction~\cite{Nagaosa2013}. These calculations are performed within the primary spatiotemporal field textures centered by hollow-circles and temporally bounded by pairs of saddle points (\cref{fig3}b). As shown in \cref{fig4}a, the toroidal excitations, illuminated by the pump polarized at 0$^\circ$ and 135$^\circ$, exhibit $N_{\mathrm{sk}}$ close to $-1$, while the other two excitations indicate values close to $+1$, in good agreement with the theoretical solutions shown by the solid lines. Moreover, $N_{\mathrm{sk}}$ is observed to reverse across both saddle points. These temporal variations in $N_{\mathrm{sk}}$ validate the ultrafast skyrmionic topological transitions within a single toroidal-pulse cycle. The $\cos(\gamma)$ values of all four toroidal excitations also reverse sign around the primary off-axis vortical field-null points at $t=0$, consistent with the theoretical values indicated by the solid lines, as shown in \cref{fig4}b. This transition of $\gamma$ reveals their key topological signature and provides further evidence of controllable spatiotemporal topological transitions.

We reconstruct the corresponding skyrmion textures at instants immediately before and after the primary off-axis vortical field-null points to further visualize the tunable topological transitions. As shown in \cref{fig4}c, the four structured fields enclosed by the skyrmion boundaries exhibit four fundamental states of N\'eel-type skyrmion textures, consistent with the measured skyrmion numbers and helicities. Collectively, the measured skyrmion number, helicity, and skyrmion textures demonstrate controllable ultrafast skyrmionic topological transitions enabled by the tunable toroidal pulses.

\section{Discussion}\label{sec5}

We have demonstrated tunable generation and comprehensive mapping of toroidal light pulses, arising from a compact, single-stage, efficient, and scalable optoelectronic metasurface. Combined patterning of local nanoantenna geometry and global metasurface texture opens a broad design space for engineering light-matter interaction symmetries and the resulting photocurrent responses. Even the lowest-order winding metasurface studied here exhibits a high degree of active tunability and striking complexity in its near- and far-field ultrafast electrodynamics. 

Beyond offering a versatile new route to toroidal light generation, the present approach introduces active control over the topology of the emitted field and direct access to its ultrafast evolution. Previous demonstrations have established the existence and characteristic structure of toroidal electromagnetic pulses, but control has largely been encoded statically into the source geometry or inferred from selected projections of the resulting fields. Here, a single device continuously transforms between magnetic-dominant and electric-dominant toroidal modes, their sign-reversed partners, and intermediate helical states through simply rotating the input linear polarization angle. Full reconstruction of the 3D electromagnetic field allows the underlying topology to be followed within a single optical cycle, directly resolving reversals of skyrmion number and helicity and the evolution among distinct skyrmionic textures. This provides an essential verification of the genuine topological structure and ultrafast evolution of toroidal pulses of light. Furthermore, the sub-cycle evolution establishes electromagnetic topology itself as an ultrafast dynamical degree of freedom, providing access to distinct topological field configurations within a single-cycle pulse.

The direct connection between vectorial currents and 3D field topology creates opportunities for structured light–matter coupling in quantum materials. Continuous control over longitudinal electric and magnetic fields can selectively address collective excitations such as phonons and magnons that are otherwise difficult to access in anisotropic systems, while toroidal fields provide natural coupling channels to toroidal dipole excitations\cite{Papasimakis2016} and dynamic anapole current configurations\cite{Savinov2019}. These nonradiating anapoles lead to a high concentration of field energy within the material and can provide an efficient means of transduction through absorption or nonlinear conversion\cite{Grinblat2016}. More generally, incorporating materials with intrinsic electronic, magnetic, or topological responses into the current-generating layer could imprint those properties directly onto the emitted field. For example, Berry-curvature-driven transverse transport\cite{Qi2011} would transform a radial photocurrent into a spiral current texture, thereby converting an intrinsic Hall response into the polarization, helicity, and electric--magnetic composition of the resulting toroidal radiation. Such material-integrated sources may therefore open a route to bidirectional transduction between electromagnetic field topology and the quantum geometry of matter.

\section{Methods}\label{sec6}

\subsection{Metasurface fabrication}

Gold metasurfaces are patterned on monolayer graphene grown via chemical vapor deposition and transferred to quartz substrates (Graphenea). Bilayer 495 A2 + 950 A2 poly(methyl methacrylate) layers are sequentially spun at 3000 rpm and soft baked at 180\,$^\circ$C for one minute, with a final layer of DisCharge added to ensure adequate charge dissipation. Electron beam patterning of the 1-mm-diameter metasurfaces is performed in a 125\,kV beam system (Elionix ELS-F125) with a 2\,nA beam current and central dose of 1700\,µC/cm$^{2}$ after proximity effect correction. This is followed by a 30\,s isopropanol (IPA) rinse to remove the DisCharge and a 60\,s development in room temperature 3:1 IPA:MIBK (methyl isobutyl ketone). A 30\,nm gold layer is applied with no adhesion layer via electron beam deposition, followed by an overnight liftoff soak in acetone. No additional processing is performed prior to loading the sample into the optical cryostat for measurements under high vacuum ($\sim$$10^{-6}$ Torr). Further details and characterization of the fabricated metasurfaces and their optical properties are provided in SI Section\,1 and Fig.\,S1.

Photocurrent readout in the Cartesian or cylindrical geometry is performed by photolithographic patterning of the graphene device geometry and gold electrodes. Following hexamethyldisilazane (HMDS) treatment to promote adhesion, AZ 5214E photoresist is spun and patterned via maskless ultraviolet writing (Heidelberg MLA 150, 375\,nm), image reversal, and development. The graphene is etched via 1\,min exposure to O$_2$ plasma (100\,W, 10\,sccm). Following the same coating, writing, and development steps, 30\,nm Au electrodes are deposited with a 3\,nm Ti adhesion layer. Photocurrent measurements are performed with 800\,nm light from a Ti:sapphire laser under continuous-wave or modelocked operation, modulated by an optical chopper and the average current read out directly through a lock-in transimpedance amplifier (SR830).

\subsection{THz hyperspectral imaging}

A schematic of the THz hyperspectral imaging system is shown in Fig.\,S2 for spatiotemporal mapping of the emitted THz radiation. We use an erbium fiber laser (Menlo Systems C-Fiber) providing 100\,fs pulses and operating at a repetition rate of 80\,MHz and a center wavelength of 1560\,nm. A fiber-coupled photoconductive antenna detector, gated by 1560\,nm pulses, is mounted on two orthogonal automated linear motion stages for 2D spatial mapping. The $E_x$ and $E_y$ field components are measured by rotating the detection axis of the photoconductive antenna. A second, synchronized and amplified free-space output at 1560\,nm is loosely focused and utilized to excite the metasurface, which is held under high vacuum in an optical cryostat (Lake Shore Janis ST-100). No discernible degradation is observed over hours or even days of exposure. The emitted THz radiation is collimated and refocused by two off-axis parabolic mirrors (diameter 2\,in, focal length 4\,in). To fully resolve the detailed field structure of the toroidal pulses despite the few-millimeter effective aperture of the photoconductive antenna silicon lens, the detection plane is offset from the focal plane by approximately 7\,cm.

\subsection{Multiphysics modeling}

We perform full-wave electrodynamics simulations with COMSOL Multiphysics 6.2 using the RF Module, combined with electronic and lattice heat flow and hydrodynamic charge flow using the Coefficient Form PDE interface. Periodic boundary conditions are utilized on the hexagonal unit cell domain. Further details on the optical, thermal, and hydrodynamic modeling are described in previous work~\cite{Pettine2024}.

\subsection {Boundary theory}

For each of the four electric-field configurations, the integration boundary was determined independently from the corresponding 2D field map. The real-valued field

\begin{equation}
\begin{aligned}
    \mathbf{E}(x,y)=\bigl[E_x(x,y),E_y(x,y),E_z(x,y)\bigr]^{\mathsf T}
\end{aligned}
\end{equation}

\noindent was sampled on a common \(401\times401\) grid spanning \(40\times40~\mathrm{mm}^2\) (which is larger than the approximately 1-inch-diameter beam size), with \(\Delta x=\Delta y=\Delta=0.1~\mathrm{mm}\). We defined the local field amplitude and normalized field direction as

\begin{equation}
\begin{aligned}
    A(x,y)
    &=
    \left[
    E_x^2(x,y)+E_y^2(x,y)+E_z^2(x,y)
    \right]^{1/2},
    \\
    \mathbf{n}(x,y)
    &=
    \frac{\mathbf{E}(x,y)}{A(x,y)}=(n_x,n_y,n_z)^{\mathsf T}.
\end{aligned}
\end{equation}

\noindent A machine-precision regularizer was included in the numerical denominator only to avoid division by zero at vanishing-amplitude pixels.

For the configurations considered here, the longitudinal component \(n_z\) provides the clearest scalar contrast between the central core and the surrounding field. We therefore constructed the segmentation map

\begin{equation}
\begin{aligned}
    \mathcal{S}(x,y)=\bigl(K_{\sigma_{\mathrm{seg}}}\ast n_z\bigr)(x,y),
    \qquad
    \sigma_{\mathrm{seg}}=0.28~\mathrm{mm}.
\end{aligned}
\end{equation}

\noindent Here, \(K_{\sigma_{\mathrm{seg}}}\) is a normalized 2D Gaussian kernel and \(\ast\) denotes 2D convolution. Thus, \(\mathcal S\) is a smoothed longitudinal-direction map used only for boundary extraction; it is not an optical-intensity distribution.

Using the polarity convention adopted for the present data, the texture center \((x_c,y_c)\) was identified as the minimum of \(\mathcal S\) among pixels with sufficiently large field amplitude. To suppress weak-field and finite-window artifacts, candidate pixels were required to lie above the 55th percentile of \(A\), and a margin equal to 12\% of the smaller image dimension was excluded along each edge. We then calculated the gradient magnitude of \(\mathcal S\), weakly smoothed it with a Gaussian kernel of standard deviation 1.2 grid spacings, and averaged it over successive annuli centered at \((x_c,y_c)\). Defining

\begin{equation}
\begin{aligned}
        r(x,y)=\sqrt{(x-x_c)^2+(y-y_c)^2},
\end{aligned}  
\end{equation}

\noindent the radius at which this annular mean was maximal was taken as the characteristic scale \(R_0\). This maximum locates the region in which the longitudinal direction changes most rapidly between the core and its surroundings. To avoid unrealistically small or boundary-dominated estimates, \(R_0\) was constrained to \(0.22D_{\min}\le R_0\le0.48D_{\min}\), where \(D_{\min}\) is the smaller image dimension and the radii are expressed in grid spacings.

A filled disk of radius \(0.95R_0\), centered at \((x_c,y_c)\), was then used to initialize a region-based Chan--Vese active contour \cite{ChanVese2001}. The contour was evolved for 220 iterations with a smoothness factor of 1.6 and a contraction bias of \(-0.05\). Because the Chan--Vese method separates the inner and outer regions according to their regional values of \(\mathcal S\), rather than following only a sharp local edge, it can retrieve a smooth, non-circular boundary from a gradually varying or distorted field texture.

After contour evolution, fixed morphological operations were used to fill holes, suppress small boundary irregularities and retain the connected component associated with the detected core. The mask was restricted to 
\begin{equation}
\begin{aligned}
    R_{\mathrm{cap}}=\min\!\left(1.65R_0,\,0.46D_{\min}\right)
\end{aligned}
\end{equation}
to prevent expansion into unrelated structures. Weak-field extensions were further removed using an amplitude gate defined by the 45th percentile of \(A\) within the local analysis window, provided that the remaining core-connected region occupied at least 1\% of that window. If no valid active-contour region remained, a core-centered circular mask was used only as a segmentation-failure safeguard and was not interpreted as an active-contour boundary.

The final core-connected binary domain for configuration \(j\) is denoted by \(\Omega_j\), with \(j=1,\ldots,4\), and its outer contour \(\partial\Omega_j\) defines the corresponding integration boundary. The initial circular mask therefore supplies only a starting estimate; the retrieved boundary is free to become non-circular. Importantly, the complete procedure is based only on the normalized longitudinal field, the local field amplitude and spatial connectivity, and the boundary is fixed before any quantity is integrated over \(\Omega_j\).

\section*{Acknowledgments}

We thank Mark Mondol and Sanghyeok Park for helpful conversations and technical support. Partial support was provided by the Los Alamos National Laboratory Laboratory Directed Research and Development (LDRD) program. This work was performed primarily at the Center for Integrated Nanotechnologies, an Office of Science User Facility operated for the U.S. Department of Energy (DOE) Office of Science. Los Alamos National Laboratory, an affirmative action equal opportunity employer, is managed by Triad National Security, LLC for the U.S. Department of Energy’s NNSA, under contract 89233218CNA000001. T.P. was supported in part by the Global University Project grant funded by the GIST in 2026. Y.L., W.Y., and Y.S. were supported by the Singapore Ministry of Education (MOE) AcRF Tier 1 (RG157/23 \& RT11/23), Singapore Agency for Science, Technology and Research (A*STAR) (M24N7c0080 \& H25-MRO3489), and Nanyang Assistant Professorship Start Up grant. M.Y. is supported by the National Science Foundation MPS-Ascend Postdoctoral Research Fellowship under Grant No. 2402151. N.G. and J.P. acknowledge support from the U.S. Department of Energy, BES DMSE, Award No. DE-FG02-08ER46521 (analysis) and Gordon and Betty Moore Foundation’s EPiQS Initiative grant GBMF9459 (manuscript writing).

\section*{Competing Interests}

The authors declare no competing interests.

\bibliography{References}

@article{Erhard2018,
  author		= "Erhard, M. and Fickler, R. and Krenn, M. and Zeilinger, A.",
  title			= "Twisted photons: New quantum perspectives in high dimensions",
  journal		= "Light Sci. Appl.",
  volume		= "7",
  pages = "17146", 
  year			= "2018"
}

@article{zhang2026skyrmions,
  title={Skyrmions based on optical anisotropy for topological encoding},
  author={Zhang, Yunqi and Wang, An Aloysius and Zhang, Runchen and Zhao, Zimo and Ma, Yifei and Liu, Ruofu and Pong, Zhi-Kai and Cai, Yuxi and He, Chao},
  journal={Light Sci. Appl.},
  volume={15},
  number={1},
  pages={254},
  year={2026},
  publisher={Nature Publishing Group UK London}
}

@article{shen2025free,
  title={Free-space topological optical textures: tutorial},
  author={Shen, Yijie and Wang, Haiwen and Fan, Shanhui},
  journal={Adv. Opt. Photonics},
  volume={17},
  number={2},
  pages={295--374},
  year={2025},
  publisher={Optica Publishing Group}
}

@article{Rosen2022,
  author		= "Quinteiro Rosen, G. F. and Tamborenea, P. I. and Kuhn, T.",
  title			= "Interplay between optical vortices and condensed matter",
  journal		= "Rev. Mod. Phys.",
  volume		= "94",
  pages = "035003",
  year			= "2022"
}

@article{Rouxel2022,
  author		= "Rouxel, J. R. and others",
  title			= "Hard {X-ray} helical dichroism of disordered molecular media",
  journal		= "Nat. Photonics",
  volume		= "16",
  pages			= "570--574",
  year			= "2022"
}

@article{Forbes2021,
  author		= "Forbes, K. A. and Andrews, D. L.",
  title			= "Orbital angular momentum of twisted light: Chirality and optical activity",
  journal		= "J. Phys. Photonics",
  volume		= "3",
  pages = "022007",
  year			= "2021"
}

@article{Sirenko2019,
  author		= "Sirenko, A. A. and others",
  title			= "Terahertz vortex beam as a spectroscopic probe of magnetic excitations",
  journal		= "Phys. Rev. Lett.",
  volume		= "122",
  pages = "237401",
  year			= "2019"
}

@article{Sirenko2021,
  author		= "Sirenko, A. A. and others",
  title			= "Total angular momentum dichroism of the terahertz vortex beams at the antiferromagnetic resonances",
  journal		= "Phys. Rev. Lett.",
  volume		= "126",
  pages			= "157401",
  year			= "2021"
}

@article{Shen2024,
  author		= "Shen, Y. and others",
  title			= "Optical skyrmions and other topological quasiparticles of light",
  journal		= "Nat. Photonics",
  volume		= "18",
  pages			= "15--25",
  year			= "2024"
}

@article{Hellwarth1996,
  author		= "Hellwarth, R. W. and Nouchi, P.",
  title			= "Focused one-cycle electromagnetic pulses",
  journal		= "Phys. Rev. E",
  volume		= "54",
  pages			= "889--895",
  year			= "1996"
}

@article{Shen2022,
  author		= "Shen, Y. and Rosales-Guzm{\'a}n, C.",
  title			= "Nonseparable states of light: From quantum to classical",
  journal		= "Laser Photonics Rev.",
  volume		= "16",
  pages = "2100533",
  year			= "2022"
}

@article{Zdagkas2020,
  author		= "Zdagkas, A. and Papasimakis, N. and Savinov, V. and Zheludev, N.",
  title			= "Space-time nonseparable pulses: Constructing isodiffracting donut pulses from plane waves and single-cycle pulses",
  journal		= "Phys. Rev. A",
  volume		= "102",
  pages			= "063512",
  year			= "2020"
}

@article{Zdagkas2022,
  author		= "Zdagkas, A. and others",
  title			= "Observation of toroidal pulses of light",
  journal		= "Nat. Photonics",
  volume		= "16",
  pages			= "523--528",
  year			= "2022"
}

@article{Forbes2025,
  author		= "Forbes, A. and Nothlawala, F. and Vall{\'e}s, A.",
  title			= "Progress in quantum structured light",
  journal		= "Nat. Photonics",
  volume		= "19",
  pages			= "1291--1300",
  year			= "2025"
}

@article{Li2012,
  author		= "Li, X. P. and Lan, T. H. and Tien, C. H. and Gu, M.",
  title			= "Three-dimensional orientation-unlimited polarization encryption by a single optically configured vectorial beam",
  journal		= "Nat. Commun.",
  volume		= "3",
  pages = "998",
  year			= "2012"
}

@article{Kasperczyk2015,
  author		= "Kasperczyk, M. and Person, S. and Ananias, D. and Carlos, L. D. and Novotny, L.",
  title			= "Excitation of magnetic dipole transitions at optical frequencies",
  journal		= "Phys. Rev. Lett.",
  volume		= "114",
  pages = "163903",
  year			= "2015"
}

@article{Papasimakis2016,
  author		= "Papasimakis, N. and Fedotov, V. A. and Savinov, V. and Raybould, T. A. and Zheludev, N. I.",
  title			= "Electromagnetic toroidal excitations in matter and free space",
  journal		= "Nat. Mater.",
  volume		= "15",
  pages			= "263--271",
  year			= "2016"
}

@article{Petrov2022,
  author		= "Petrov, N. V. and Sokolenko, B. and Kulya, M. S. and Gorodetsky, A. and Chernykh, A. V.",
  title			= "Design of broadband terahertz vector and vortex beams: I. {Review} of materials and components",
  journal		= "Light Adv. Manuf.",
  volume		= "3",
  pages			= "43",
  year			= "2022"
}

@article{Imai2012,
  author		= "Imai, R. and others",
  title			= "Terahertz vector beam generation using segmented nonlinear optical crystals with threefold rotational symmetry",
  journal		= "Opt. Express",
  volume		= "20",
  pages			= "21896--21904",
  year			= "2012"
}

@article{Deveikis2022,
  author		= "Deveikis, J. and Lloyd-Hughes, J.",
  title			= "Multi-pixel photoconductive emitters for the controllable generation of azimuthal and radial terahertz beams",
  journal		= "Opt. Express",
  volume		= "30",
  pages			= "43293--43300",
  year			= "2022"
}

@article{Imai2014,
  author		= "Imai, R. and Kanda, N. and Higuchi, T. and Konishi, K. and Kuwata-Gonokami, M.",
  title			= "Generation of broadband terahertz vortex beams",
  journal		= "Opt. Lett.",
  volume		= "39",
  pages			= "3714--3717",
  year			= "2014"
}

@article{Wang2024,
  author		= "Wang, S. J. and others",
  title			= "Flexible generation of structured terahertz fields via programmable exchange-biased spintronic emitters",
  journal		= "eLight",
  volume		= "4",
  pages = "11",
  year			= "2024"
}

@article{Jana2024,
  author		= "Jana, K. and others",
  title			= "Quantum control of flying doughnut terahertz pulses",
  journal		= "Sci. Adv.",
  volume		= "10",
  pages = "eadl1803",
  year			= "2024"
}

@article{Niu2026,
  author		= "Niu, L. and others",
  title			= "Electric-magnetic-switchable free-space skyrmions in toroidal light pulses via a nonlinear metasurface",
  journal		= "Optica",
  volume		= "13",
  pages			= "203--210",
  year			= "2026"
}

@article{Pettine2023,
  author		= "Pettine, J. and others",
  title			= "Ultrafast terahertz emission from emerging symmetry-broken materials",
  journal		= "Light Sci. Appl.",
  volume		= "12",
  pages			= "133",
  year			= "2023"
}

@article{Pettine2024,
  author		= "Pettine, J. and others",
  title			= "Light-driven nanoscale vectorial currents",
  journal		= "Nature",
  volume		= "626",
  pages			= "984--989",
  year			= "2024"
}

@article{Wei2020,
  author		= "Wei, J. X. and others",
  title			= "Zero-bias mid-infrared graphene photodetectors with bulk photoresponse and calibration-free polarization detection",
  journal		= "Nat. Commun.",
  volume		= "11",
  pages			= "6404",
  year			= "2020"
}

@book{Powell2010,
  author		= "Powell, R. C.",
  title			= "Symmetry, group theory, and the physical properties of crystals",
  publisher    = "Springer",
  year			= "2010"
}

@article{Taghinejad2023,
  author		= "Taghinejad, M. and others",
  title			= "Determining hot-carrier transport dynamics from terahertz emission",
  journal		= "Science",
  volume		= "382",
  pages			= "299--305",
  year			= "2023"
}

@article{Matsubara2022,
  author		= "Matsubara, M. and others",
  title			= "Polarization-controlled tunable directional spin-driven photocurrents in a magnetic metamaterial with threefold rotational symmetry",
  journal		= "Nat. Commun.",
  volume		= "13",
  pages = "6708",
  year			= "2022"
}

@article{McDonnell2022,
  author		= "McDonnell, C. and Deng, J. H. and Sideris, S. and Li, G. X. and Ellenbogen, T.",
  title			= "Terahertz metagrating emitters with beam steering and full linear polarization control",
  journal		= "Nano Lett.",
  volume		= "22",
  pages			= "2603--2610",
  year			= "2022"
}

@article{Wang2024APR,
  author		= "Wang, R. and others",
  title			= "Observation of resilient propagation and free-space skyrmions in toroidal electromagnetic pulses",
  journal		= "Appl. Phys. Rev.",
  volume		= "11",
  number		= "3",
  pages			= "031411",
  year			= "2024"
}

@article{Wang2025SciAdv,
  author		= "Wang, R. and others",
  title			= "Hybrid electromagnetic toroidal vortices",
  journal		= "Sci. Adv.",
  volume		= "11",
  number		= "8",
  pages			= "eads4797",
  year			= "2025"
}

@article{Shen2021,
  author		= "Shen, Y. and others",
  title			= "Supertoroidal light pulses as electromagnetic skyrmions propagating in free space",
  journal		= "Nat. Commun.",
  volume		= "12",
  pages			= "5891",
  year = "2021"
}

@article{Shen2024NDSTP,
  author		= "Shen, Y. and others",
  title			= "Nondiffracting supertoroidal pulses and optical “{Kármán} vortex streets”",
  journal		= "Nat. Commun.",
  volume		= "15",
  pages			= "4863",
  year			= "2024"
}

@article{Shen2026STSO,
  author		= "Shen, Y. and others",
  title			= "Space-time superoscillations",
  journal		= "Nat. Commun.",
  volume		= "17",
  pages			= "2053",
  year			= "2026"
}

@article{Shi2026,
  author		= "Shi, S. and others",
  title			= "Toroidal helical pulses",
  journal		= "Commun. Phys.",
  volume		= "9",
  pages			= "287",
  year			= "2026"
}

@article{cheng2025metrology,
  title={Metrology with a twist: probing and sensing with vortex light},
  author={Cheng, Mingjian and Jiang, Wenjie and Guo, Lixin and Li, Jiangting and Forbes, Andrew},
  journal={Light Sci. Appl.},
  volume={14},
  number={1},
  pages={4},
  year={2025},
  publisher={Nature Publishing Group UK London}
}

@article{he2022towards,
  title={Towards higher-dimensional structured light},
  author={He, Chao and Shen, Yijie and Forbes, Andrew},
  journal={Light Sci. Appl.},
  volume={11},
  number={1},
  pages={205},
  year={2022},
  publisher={Nature Publishing Group UK London}
}

@article{shen2019optical,
  title={Optical vortices 30 years on: {OAM} manipulation from topological charge to multiple singularities},
  author={Shen, Yijie and Wang, Xuejiao and Xie, Zhenwei and Min, Changjun and Fu, Xing and Liu, Qiang and Gong, Mali and Yuan, Xiaocong},
  journal={Light Sci. Appl.},
  volume={8},
  number={1},
  pages={90},
  year={2019},
  publisher={Nature Publishing Group UK London}
}

@article{vignjevic2026state,
  title={State tomography of toroidal light pulses},
  author={Vignjevic, Luka and Shen, Yijie and Papasimakis, Nikitas and Zheludev, Nikolay I},
  journal={J. Opt.},
  volume={28},
  number={2},
  pages={025601},
  year={2026},
  publisher={IOP Publishing}
}

@article{Savinov2019,
  author		= "Savinov, V. and Papasimakis, N. and Tsai, D. P. and Zheludev, N. I.",
  title			= "Optical anapoles",
  journal		= "Commun. Phys.",
  volume		= "2",
  pages			= "69",
  year			= "2019"
}

@article{Grinblat2016,
  author		= "Grinblat, G. and Li, Y. and Nielsen, M. P. and Oulton, R. F. and Maier, S. A.",
  title			= "Enhanced third harmonic generation in single germanium nanodisks excited at the anapole mode",
  journal		= "Nano Lett.",
  volume		= "16",
  pages			= "4635-4640",
  year			= "2016"
}

@article{Nagaosa2013,
  author    = "Nagaosa, N. and Tokura, Y.",
  title     = "Topological properties and dynamics of magnetic skyrmions",
  journal   = "Nat. Nanotechnol.",
  volume    = "8",
  pages     = "899--911",
  year      = "2013"
}

@article{Qi2011,
  author    = "Qi, X.-L. and Zhang, S.-C.",
  title     = "Topological insulators and superconductors",
  journal   = "Rev. Mod. Phys.",
  volume    = "83",
  pages     = "1057--1110",
  year      = "2011"
}

@article{Wang2024LSA,
  author    = "Wang, A. A. and others",
  title     = "Topological protection of optical skyrmions
through complex media",
  journal   = "Light Sci. Appl.",
  volume    = "13",
  pages     = "314",
  year      = "2024"
}

@article{Shautsova2018NatCommun,
  author    = "Shautsova, V. and others",
  title     = "Plasmon induced thermoelectric effect in graphene",
  journal   = "Nat. Commun.",
  volume    = "9",
  pages     = "5190",
  year      = "2018"
}

@ARTICLE{ChanVese2001,
  author={Chan, T.F. and Vese, L.A.},
  journal={IEEE Trans. Image Process.}, 
  title={Active contours without edges}, 
  year={2001},
  volume={10},
  number={2},
  pages={266-277},
  doi={10.1109/83.902291}}

@article{Elias2009Science,
    author = {D. C. Elias  and others},
    title = {Control of Graphene's Properties by Reversible Hydrogenation: Evidence for Graphane},
    journal = {Science},
    volume = {323},
    number = {5914},
    pages = {610-613},
    year = {2009},
    doi = {10.1126/science.1167130}
}

@article{Luo2014NatCommun,
    author = {L. Luo  and others},
    title = {Broadband terahertz generation from metamaterials},
    journal = {Nat. Commun.},
    volume = {5},
    pages = {3055},
    year = {2014},
    doi = {10.1038/ncomms4055}
}

@article{Guo2026NatCommun,
  author    = "Guo, Z. and others",
  title     = "Topological robustness of classical and quantum optical skyrmions in atmospheric turbulence",
  journal   = "Nat. Commun.",
  volume    = "17",
  pages     = "2085",
  year      = "2026"
}

@article{Chen2026NatElectron,
  author    = "Chen, L. and others",
  title     = "Programmable skyrmions for communication and sensing",
  journal   = "Nat. Electron.",
  volume    = "9",
  pages     = "479--488",
  year      = "2026"
}

@article{Kam2026eLight,
  author    = "Kam, A. and others",
  title     = "Quantum skyrmions and high dimensional entanglement mediated by nanophotonics",
  journal   = "eLight",
  volume    = "6",
  pages     = "13",
  year      = "2026"
}

\end{document}


\title{
  {\normalsize\bfseries Supplementary Information for}\\[0.4em]
  {Topological terahertz fields from tunable vectorial currents}
}

\author{Jacob Pettine}
\thanks{These authors contributed equally to this work.}
\affiliation{Center for Integrated Nanotechnologies, Los Alamos National Laboratory, Los Alamos, New Mexico 87545, USA}
\affiliation{Department of Physics, Massachusetts Institute of Technology, Cambridge, Massachusetts 02139, USA}

\author{Tae Gwan Park}
\thanks{These authors contributed equally to this work.}
\affiliation{Center for Integrated Nanotechnologies, Los Alamos National Laboratory, Los Alamos, New Mexico 87545, USA}
\affiliation{Department of Physics and Photon Science, Gwangju Institute of Science and Technology, Gwangju 61005, Republic of Korea}

\author{Yanhe Lyu}
\affiliation{Centre for Disruptive Photonic Technologies, School of Physical and Mathematical Sciences, Nanyang Technological University, Singapore 637371, Singapore}

\author{Wangke Yu}
\affiliation{Centre for Disruptive Photonic Technologies, School of Physical and Mathematical Sciences, Nanyang Technological University, Singapore 637371, Singapore}

\author{John B. Maier}
\affiliation{Department of Physics, Massachusetts Institute of Technology, Cambridge, Massachusetts 02139, USA}

\author{Matthew Yeung}
\affiliation{Department of Physics, Massachusetts Institute of Technology, Cambridge, Massachusetts 02139, USA}

\author{Rujuta Sane}
\affiliation{Department of Physics, Massachusetts Institute of Technology, Cambridge, Massachusetts 02139, USA}

\author{Chun-Chieh Chang}
\affiliation{Center for Integrated Nanotechnologies, Los Alamos National Laboratory, Los Alamos, New Mexico 87545, USA}

\author{Nuh Gedik}
\affiliation{Department of Physics, Massachusetts Institute of Technology, Cambridge, Massachusetts 02139, USA}

\author{Yijie Shen}\email{yijie.shen@ntu.edu.sg}
\affiliation{Centre for Disruptive Photonic Technologies, School of Physical and Mathematical Sciences, Nanyang Technological University, Singapore 637371, Singapore}
\affiliation{School of Electrical and Electronics Engineering, Nanyang Technological University, Singapore 639798, Singapore}

\author{Hou-Tong Chen}\email{chenht@lanl.gov}
\affiliation{Center for Integrated Nanotechnologies, Los Alamos National Laboratory, Los Alamos, New Mexico 87545, USA}

\maketitle
\tableofcontents
\clearpage

\section{S1. Metasurface characterization}\label{sec1}

\begin{figure*}[h]
    \centering
    \includegraphics[width=0.95\textwidth]{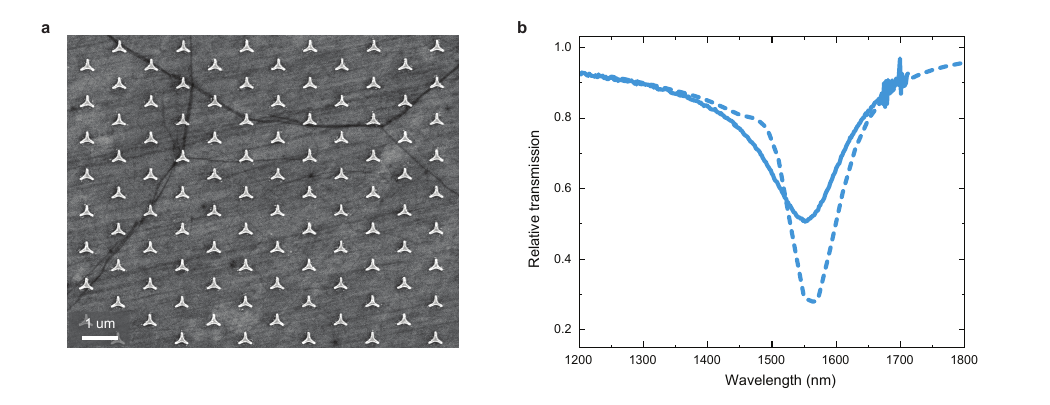}
    \caption{\textbf{Metasurface characteristics.}
    \textbf{a}, Scanning electron micrograph of a uniform three-petal metasurface on graphene. 
    \textbf{b}, Measured (solid) and simulated (dashed) relative transmission spectra showing the plasmonic resonance dip designed around 1560\,nm.}
    \label{FigS_TP1550b}
\end{figure*}

A representative scanning electron micrograph of a uniform gold three-petal metasurface fabricated with a high degree of regularity across a millimeter-scale region on graphene is shown in \cref{FigS_TP1550b}a. The graphene is grown via chemical vapor deposition (CVD) and transferred to quartz. The microscopic scale darker lines are multilayer folds of graphene and have negligible effect on the nanoscopic optical drive or on the micro-to-macroscopic photocurrent responses. 

The relative transmission spectrum through the graphene/quartz chip with the metasurface referenced to a region without the metasurface is shown in \cref{FigS_TP1550b}b (solid line). The 1560\,nm plasmonic resonance dip is in good agreement with the response calculated via finite element simulation (dashed line), with a small blue shift due to slight truncation of the three-petal arms attributed to resolution limitations in the fabrication. Due to the $C_3$ structural symmetry, the linear transmission response is independent of the incident light polarization. The characteristic broadening of the measured spectrum is due to residual structural inhomogeneity along with additional bulk and interface damping in the polycrystalline gold. Similar optical responses exist in the winding metasurfaces, which exhibit the same local array structure despite globally curved patterning.

\clearpage

\section{S2. Terahertz emission spectroscopy}\label{sec2}

\begin{figure*}[h!]
    \centering
    \includegraphics[width=0.8\textwidth]{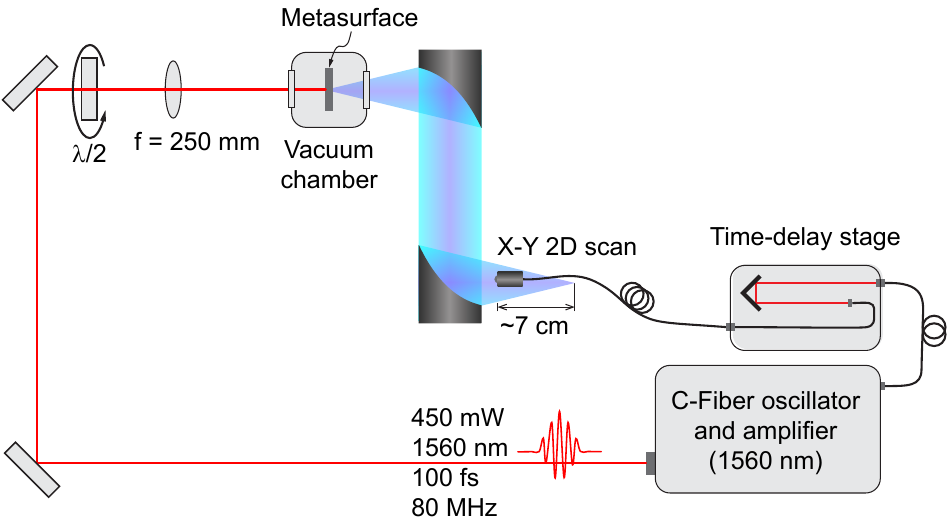}
    \caption{\textbf{Schematic of THz hyperspectral imaging setup.}
    Femtosecond pulses generated by a 1560 nm fiber laser are loosely focused onto the 1-mm-diameter winding metasurface, producing THz vector pulses that are collimated and refocused by two off-axis parabolic mirrors. The focusing THz fields are mapped with a scanning photoconductive antenna (PCA) detector between the parabolic mirror and the focal plane for optimal resolution. Further details provided in the Methods and supplementary text.}
    \label{FigS_ExperimentSchematic}
\end{figure*}

\begin{figure*}[h!]
    \centering
    \includegraphics[width=\textwidth]{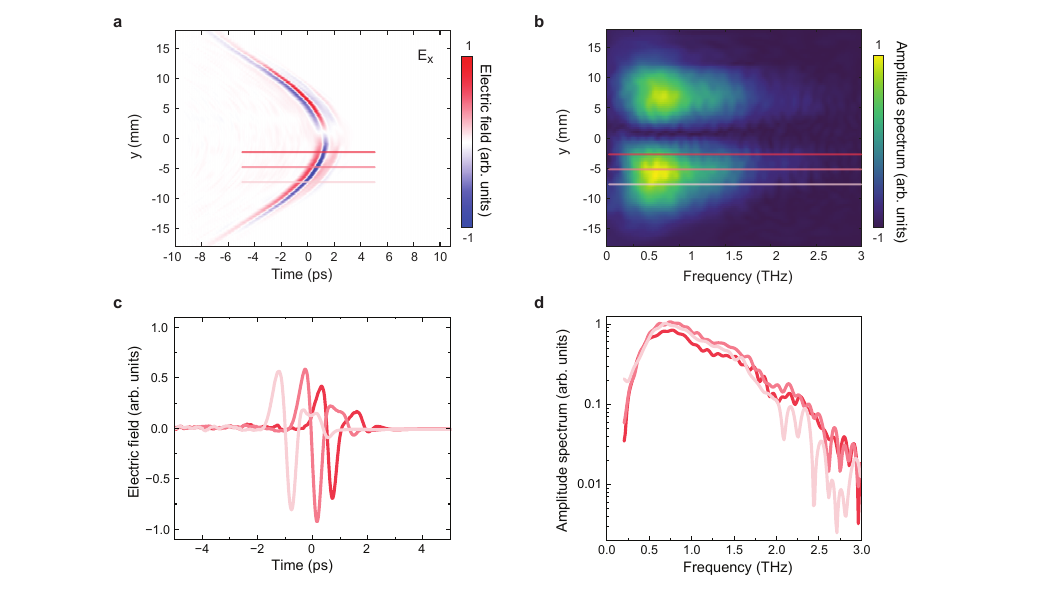}
    \caption{\textbf{Spectral properties of THz pulses.}
    \textbf{a}, Spatial map of THz time traces measured between refocusing off-axis parabolic mirror and focal plane for representative 90$^\circ$ incident polarization (azimuthal photocurrent) dataset. \textbf{b}, Spatially-resolved fast-Fourier transform. \textbf{c}, Time and \textbf{d}, frequency slices at different spatial positions along the $y$ axis, showing similar spectral properties and broad support.}
    \label{FigS_Spectra90deg}
\end{figure*}

The optical setup for THz emission spectroscopy and hyperspectral imaging is schematically illustrated in \cref{FigS_ExperimentSchematic}. Given measurement away from the focal plane of the refocusing off-axis parabolic mirror to optimize spatial resolution of the hyperspectral imaging, the raw THz spatial maps exhibit a high degree of curvature. This is seen in \cref{FigS_Spectra90deg}a for the $E_x$ component of the 90$^\circ$ (azimuthal photocurrent) dataset, along the $y$ axis. To analyze the toroidal properties of the pulses we either apply (i) a fitted spherical phase correction or (ii) numerical propagation to the focal plane, both of which are discussed further in the main text and below. Here we evaluate the raw temporal and spectral properties of the measured pulse without any corrections except mild windowing to remove spurious signal at late times and interpolation plus zero padding to improve the quality of the fast-Fourier transform, shown in \cref{FigS_Spectra90deg}b. Lineouts across the temporal and spectral profiles are shown in \cref{FigS_Spectra90deg}c and \cref{FigS_Spectra90deg}d, respectively, revealing spatially-varying amplitude (and uncorrected phase) but only minor changes in the spectra consistent with a weakly visible space-time coupling. This isodiffraction behavior is clarified near the focal plane, discussed below. The key observation in this dataset is the broad spectral support of the toroidal THz pulses. While measured beyond 3\,THz, this remains limited by the signal/noise ratio of the measurement and the photoconductive antenna response because of the large THz beam diameter (slightly larger than 1\,in). With broadband detection, significant spectral content is expected out beyond 4\,THz given the $\sim$250\,fs simulated photothermoelectric drive\cite{Pettine2024}. Broader spectra may be expected for faster optical excitation pulses with less total energy deposited, along with faster momentum-relaxing scattering rates in lower-mobility graphene---e.g., with the addition of defects due to hydrogenation\cite{Elias2009Science}.

\clearpage

\clearpage

\section{S3. Unbounded vectorial currents and near-field distributions}\label{sec3}

Optoelectronic metasurfaces represent a unique type of unbounded 2D circuit, with each nanoantenna establishing a local electromotive force. Here, this is the result of unequal diffusion away from the nanoantenna tip hot spots into the gold-doped and bare graphene. The net force at short times is effectively the vectorial sum of each local tip hot spot contribution, $\textbf{f}_\mathrm{emf}\propto\Delta S \Delta T_e$, where $\Delta S = S_\mathrm{doped}-S_\mathrm{bare}$ is the Seebeck coefficient difference and $\Delta T_e = T_e-T_0$ is the electronic temperature increase relative to equilibrium. In simple directional arrays with nanoantennas driving currents between two electrodes, it has been shown that increasing the number of nanoantenna columns between the electrodes increases the photocurrent proportionately\cite{Wei2020}. This behaves as a Thevenin equivalent circuit for a series of voltage sources, with the same overall resistance between the two electrodes. However, the increase of current with increasing row number as well points to behaviors that are simultaneously characteristic of conventional voltage and current sources.

To visualize the ultrafast current flow across the $R=500$\,µm winding metasurfaces, we solve the time-domain Drude equation numerically for the surface current density, $\textbf{j}(\textbf{r},t)$,
\begin{equation}
    \tau_\textrm{D} \frac{\partial\textbf{j}(\textbf{r},t)}{\partial t} + \textbf{j}(\textbf{r},t) = \sigma_\textrm{dc} \textbf{E}_\textrm{emf}(\textbf{r},t),
\end{equation}
along with the continuity equation
\begin{equation}
    \nabla\cdot \textbf{j}(\textbf{r},t) = -\frac{\partial \rho(\textbf{r},t)}{\partial t}.
\end{equation}
Here $\rho(\textbf{r},t)$ is the transient charge density, $\tau_\textrm{D} = \mu E_\textrm{F}/(ev_\textrm{F}^2)\approx 50$\,fs is the momentum relaxation time due to inelastic scattering and $\sigma_\textrm{dc} = e\mu E_\textrm{F}^2/(\pi\hbar^2 v_\textrm{F}^2) \approx 2.2$\,mS is the dc conductivity for CVD graphene mobility $\mu \approx$ 1500 $\textrm{cm}^2/\textrm{V}\textrm{s}$, environmentally-doped $E_\textrm{F} \approx -350$\,meV, and Fermi velocity $v_\textrm{F} = 10^6$\,m/s. The source field,
\begin{equation}
    \textbf{E}_\textrm{emf} = \textbf{E}_\textrm{PTE} + \textbf{E}_q
\end{equation}
consists of the photothermoelectric transient metasurface drive, 
\[
    \textbf{E}_\textrm{PTE}(\textbf{r},t) = E_\textrm{PTE,0}(\textbf{r},t)\left[\sin(2\theta) \hat{\boldsymbol{r}} + \cos(2\theta)\hat{\boldsymbol{\theta}}\right],
\]
and the opposing radial field due to charging from compressional mode contributions, $\textbf{E}_q(\textbf{r},t) = E_q(t)\hat{\boldsymbol{r}}$. Due to the uniform metasurface density and corresponding electromotive force field, the short-time charge buildup progressively increases toward the center of the metasurface $\rho \propto r^{-1}$ until plateauing within the nanoantenna-free core region, several micrometers in diameter. To set the magnitude of $\textbf{j}$ and thus $E_\textrm{emf}$, we approximate a damage threshold one thousand-fold higher than the peak current around 800\,µA inferred from the measured current of 10\,nA, the 80\,MHz repetition rate, and a $\sim$250\,fs decay time for the photocurrent determined from multiphysics simulations\cite{Pettine2024} (accounting for the full electron and lattice thermodynamic evolution of the force field). Thus $j_r(\textbf{r},t)=I_r(\textbf{r},t)/2\pi r$ or $j_\theta(\textbf{r},t)=I_\theta(\textbf{r},t)/R$ offer a quantitative lower bound on the magnitude of the response, which may indeed be stronger due to loss and limited bandwidth of the lock-in readout.

The near-field \textbf{E} and \textbf{B} distributions are determined from the full time-evolving photocurrent and charge distributions. The $\rho(\textbf{r},t)$ and $\textbf{j}(\textbf{r},t)$ are Fourier transformed in time and decomposed in a continuous basis of cylindrically-symmetric zeroth- and first-order Bessel functions of the first kind, respectively, via Hankel transforms. Maxwell's equations applied to each of these terms yield
\begin{equation}
\begin{split}
    E_r(k,\omega,z=0)=\frac{i k_{z,u}k_{z,d}}{\epsilon_0(\epsilon_d k_{z,u} + \epsilon_u k_{z,d})k}\rho(k,\omega), \\    
    E_\theta(k,\omega,z=0)=-\frac{\mu_0\omega }{k_{z,u} + k_{z,d}}j_\theta (k,\omega), \\ 
    E_z(k,\omega,z=0)=\frac{k_{z,d}}{\epsilon_0(\epsilon_d k_{z,u} + \epsilon_u k_{z,d})}\rho(k,\omega), \\
    B_r(k,\omega,z=0) = \frac{\mu_0 k_{z,u} }{k_{z,u} + k_{z,d}}j_\theta(k,\omega), \\
    B_\theta(k,\omega,z=0) = -\frac{\mu_0 \epsilon_u k_{z,d} }{\epsilon_d k_{z,u} + \epsilon_u k_{z,d}}j_r(k,\omega), \\
    B_z(k,\omega,z=0) = -\frac{i \mu_0 k }{k_{z,u} + k_{z,d}}j_\theta(k,\omega),
\end{split}
\end{equation}
where $k$ is the in-plane wavevector magnitude and
\begin{equation}
    k_{z,i} = \sqrt{\epsilon_i \left( \frac{\omega}{c}\right)^2-k^2},
\end{equation}
for the upper ($i = u$) air and lower ($i=d$) SiO$_2$ media. Propagation to finite $z$ then simply involves multiplying each component by $e^{-i k_{z,u}z}$. The spatial and temporal field evolution is recovered via inverse Hankel and Fourier transforms, respectively. Nonradiative evanescent fields can be explicitly removed by excluding imaginary $k_z$ components, where $k > \sqrt{\epsilon_u}\omega/c$. Such exclusion leads to a reduction in field amplitudes by approximately a factor of two.

\begin{figure*}[h!]
    \centering
    \includegraphics[width=\textwidth]{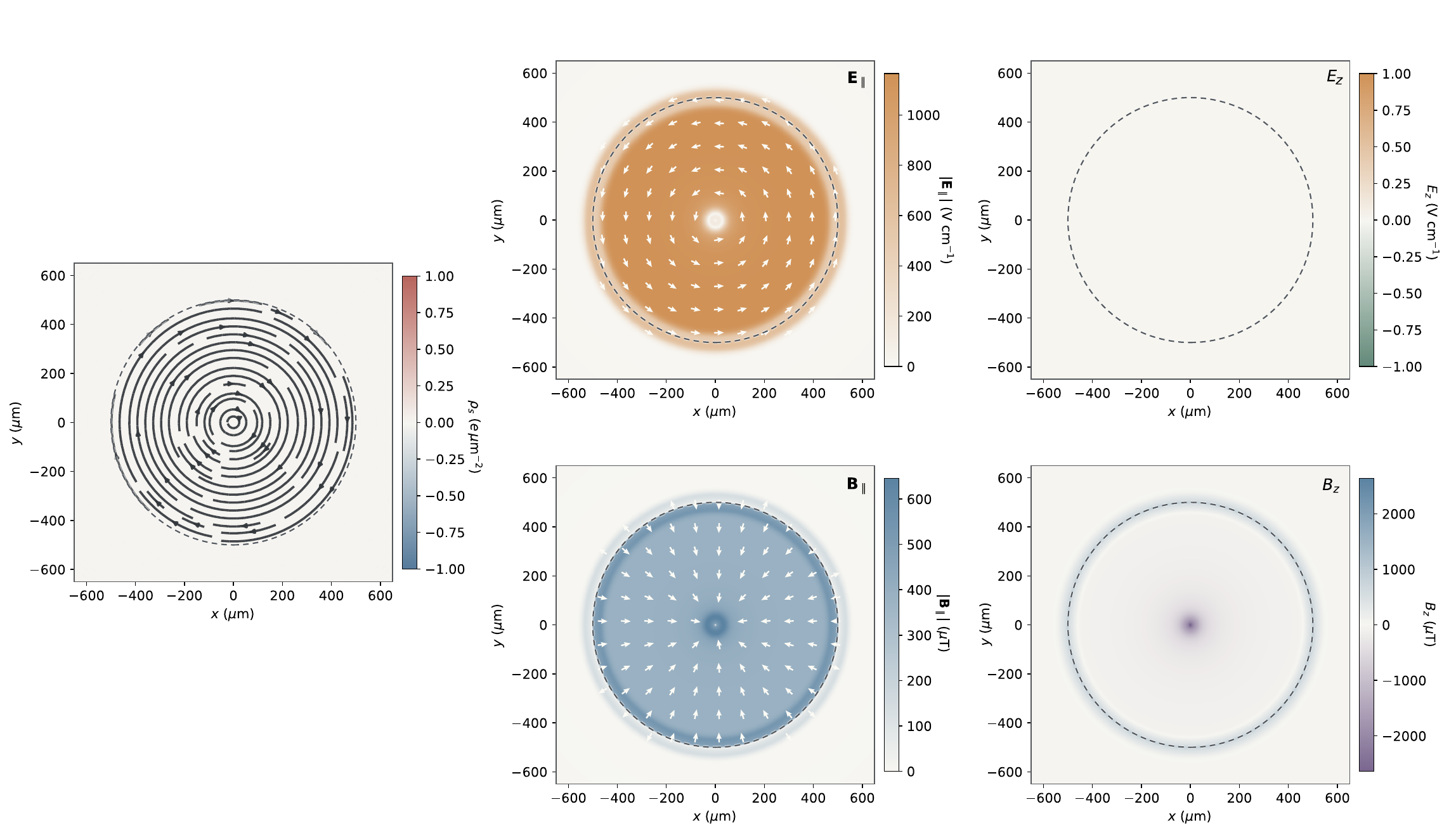}
    \caption{\textbf{Simulated circulating ultrafast current and near field distributions at 90$^\circ$ incident polarization.}
    Two-dimensional photocurrent flow lines for the clockwise-azimuthal circulating mode (left) yield zero charge buildup and thus zero $E_z$, while strong $\sim$2\,mT magnetic near-fields are estimated semi-quantitatively 200\,fs after excitation 1\,µm above the central metasurface region.}
    \label{FigS_ModelPC90deg}
\end{figure*}

\begin{figure*}[h!]
    \centering
    \includegraphics[width=\textwidth]{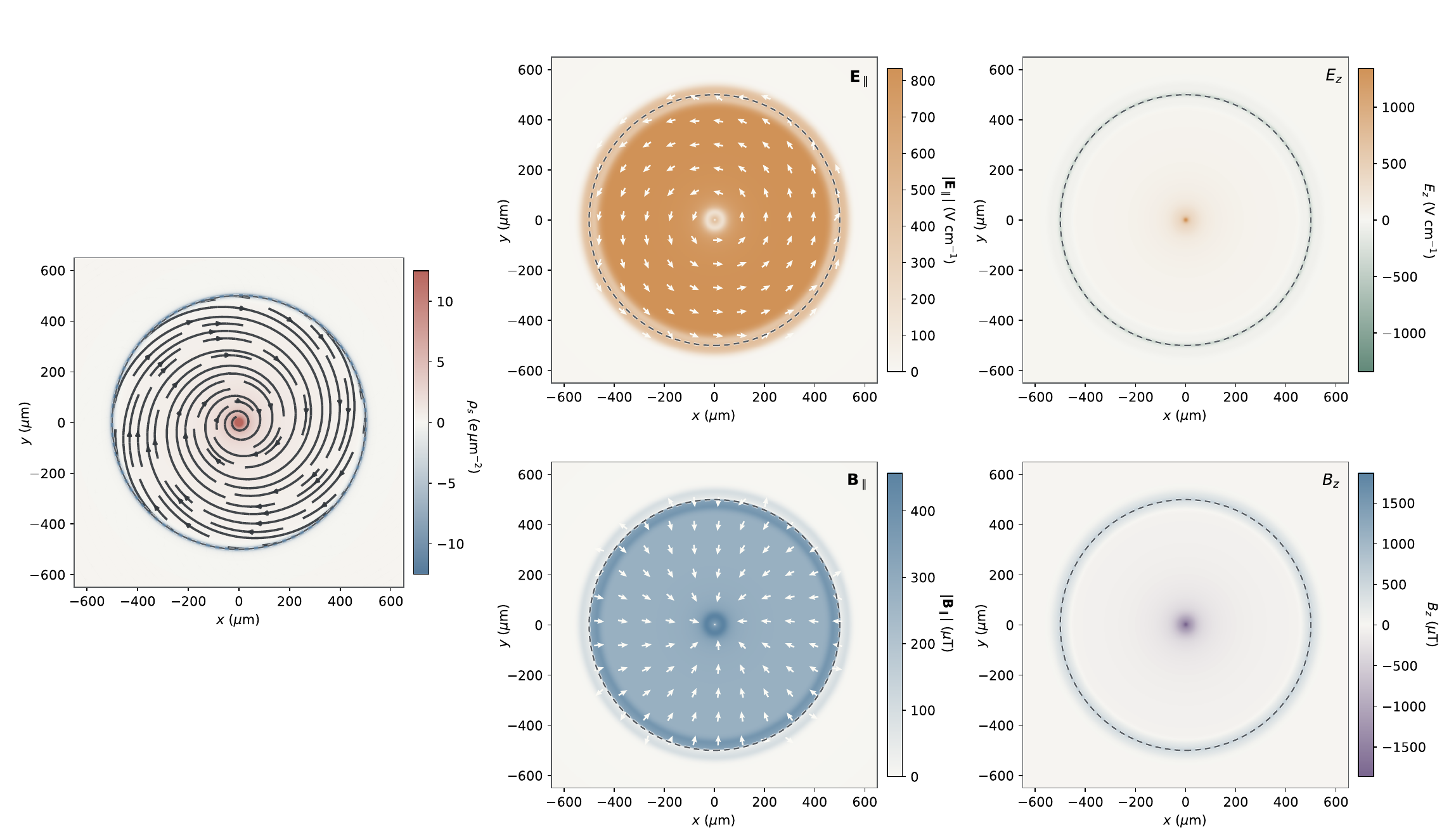}
    \caption{\textbf{Simulated spiral ultrafast current and near field distributions at 112.5$^\circ$ incident polarization.}
    Two-dimensional photocurrent flow lines for the intermediate mode (left) yield finite charge buildup and nonzero $E_z$ and $B_z$, shown 200\,fs after excitation 1\,µm above the central metasurface region.}
    \label{FigS_ModelPC112.5deg}
\end{figure*}

\begin{figure*}[h!]
    \centering
    \includegraphics[width=\textwidth]{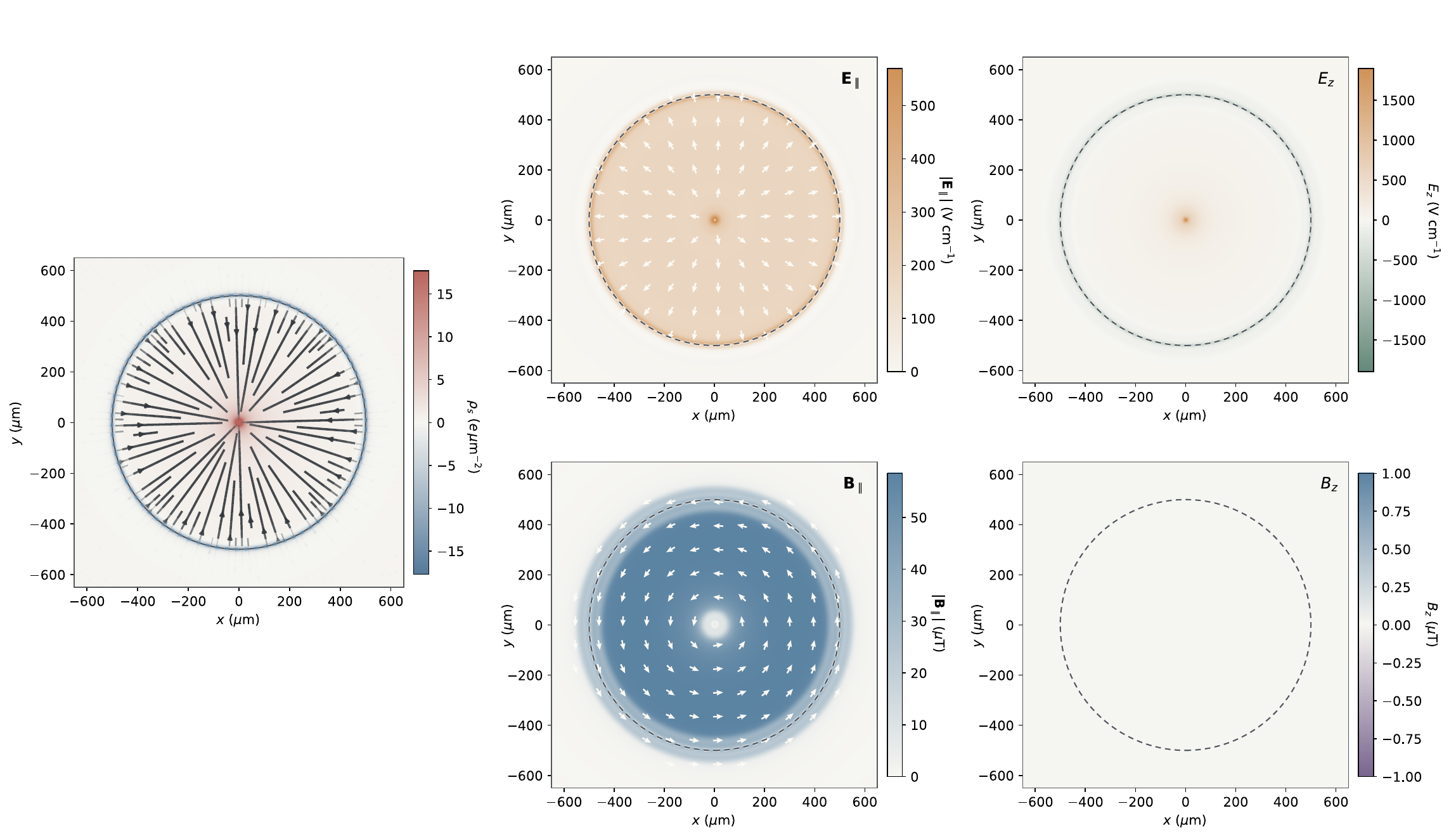}
    \caption{\textbf{Simulated compressional ultrafast current and near field distributions at 135$^\circ$ incident polarization.}
    Two-dimensional photocurrent flow lines for the inward-radial compressional mode (left) yield zero vorticity and thus zero $B_z$, while strong $\sim$1.5\,kV/cm electric near-fields are estimated semi-quantitatively 200\,fs after excitation 1\,µm above the central metasurface region.}
    \label{FigS_ModelPC135deg}
\end{figure*}

\begin{figure*}[h!]
    \centering
    \includegraphics[width=\textwidth]{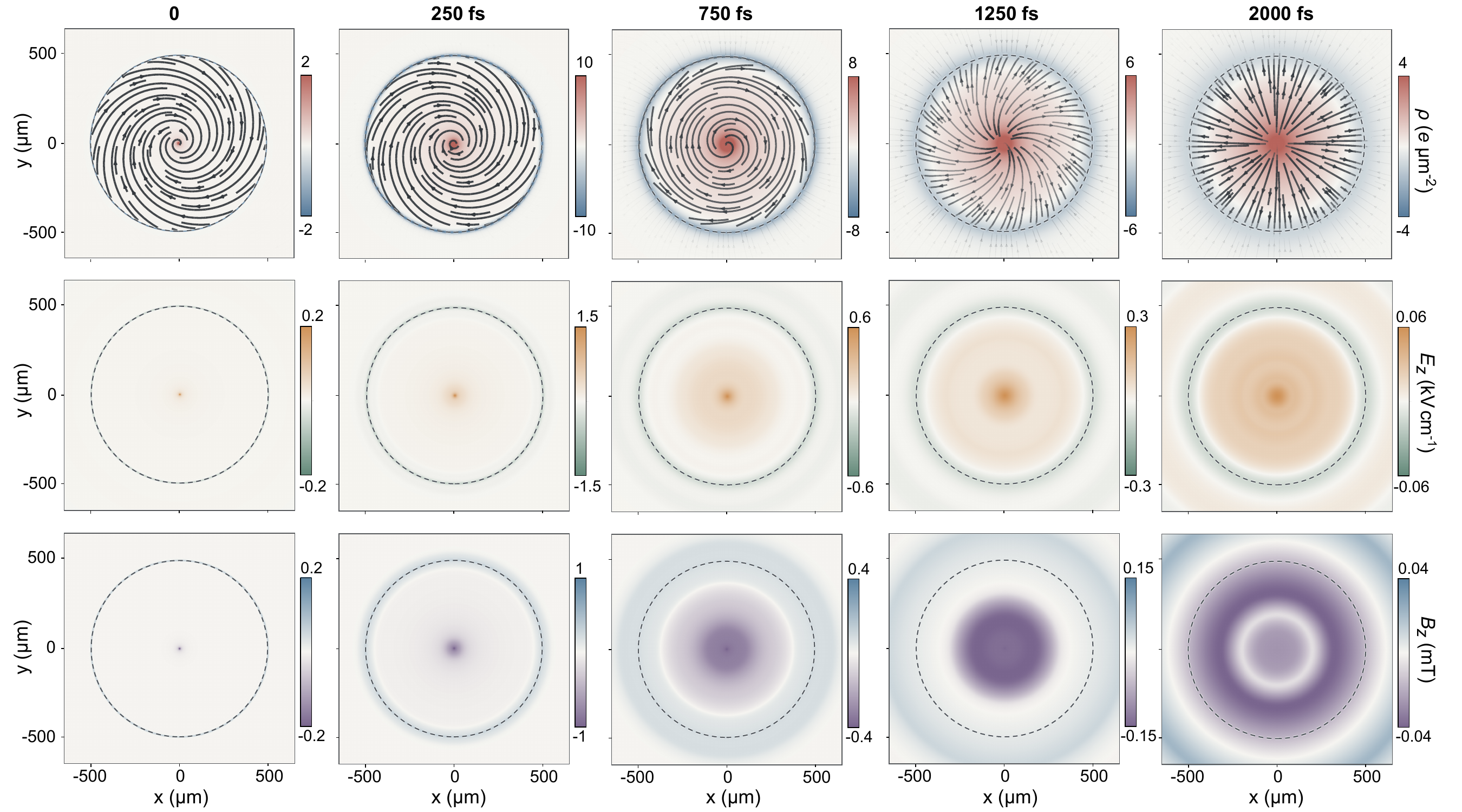}
    \caption{\textbf{Ultrafast photocurrent, charge, and field evolution at 120$^\circ$ incident polarization.}
    Flow lines (top panels) show spiral current and corresponding central concentration of charge and fields at short timescales. After several hundred femtoseconds, the driving electromotive force has decayed and the restoring Coulomb force of the charge field eventually reverses the current spiral polarity. By 1.25\,ps the restoring radial current is dominant. Flow lines and colormaps are normalized to peak value, though it should be noted (not shown explicitly) that current density decays quickly after 250\,fs as the dominant driving photothermoelectric force is decayed.}
    \label{FigS_ModelPCtdep}
\end{figure*}

Snapshots of the photocurrent, charge, and field distributions are shown 1\,µm above the metasurface in \cref{FigS_ModelPC90deg} (90$^\circ$ incident polarization), \cref{FigS_ModelPC112.5deg} (112.5$^\circ$), and \cref{FigS_ModelPC135deg} (135$^\circ$) 200\,fs after excitation. Retardation of the in-driving electromotive field---which propagates at roughly $c/\sqrt{2.5}$ or $\sim$190\,µm/ps given the effective $\epsilon\approx 2.5$ of the surrounding medium---is not explicitly accounted for in the simulations (while retardation of the free-space fields is), though it remains minimal on the few-hundred femtosecond timescales over which the driving PTE field persists. On longer timescales, the restoring field due to charge leads to an eventual reversal of the radial current shown in \cref{FigS_ModelPCtdep}. 

The modeled charge flow and fields clarify the skyrmion topology already encoded directly within the photocurrents and near fields. In the two-dimensional photocurrents, the non-existent $z$ component is replaced by the photocurrent divergence (or charge) for the compressional mode and the photocurrent vorticity for the circulating mode. These map directly onto the $(E_x,E_y,E_z)$ and $(B_x,B_y,B_z)$ near-field distributions, respectively. While the built-in optoelectronic skyrmion topology of the patterned ultrafast photocurrents may prove to be of value, we leave this as a subject for further investigation. Meanwhile, we see that the near-field topology is directly inherited by the nonparaxial diverging or refocused far-field pulses. 

\clearpage

\section{S4. Component-resolved spatiotemporal fields of the tunable toroidal states}
\label{sec4}

\begin{figure*}[t]
    \centering
    \includegraphics[width=0.95\textwidth]
    {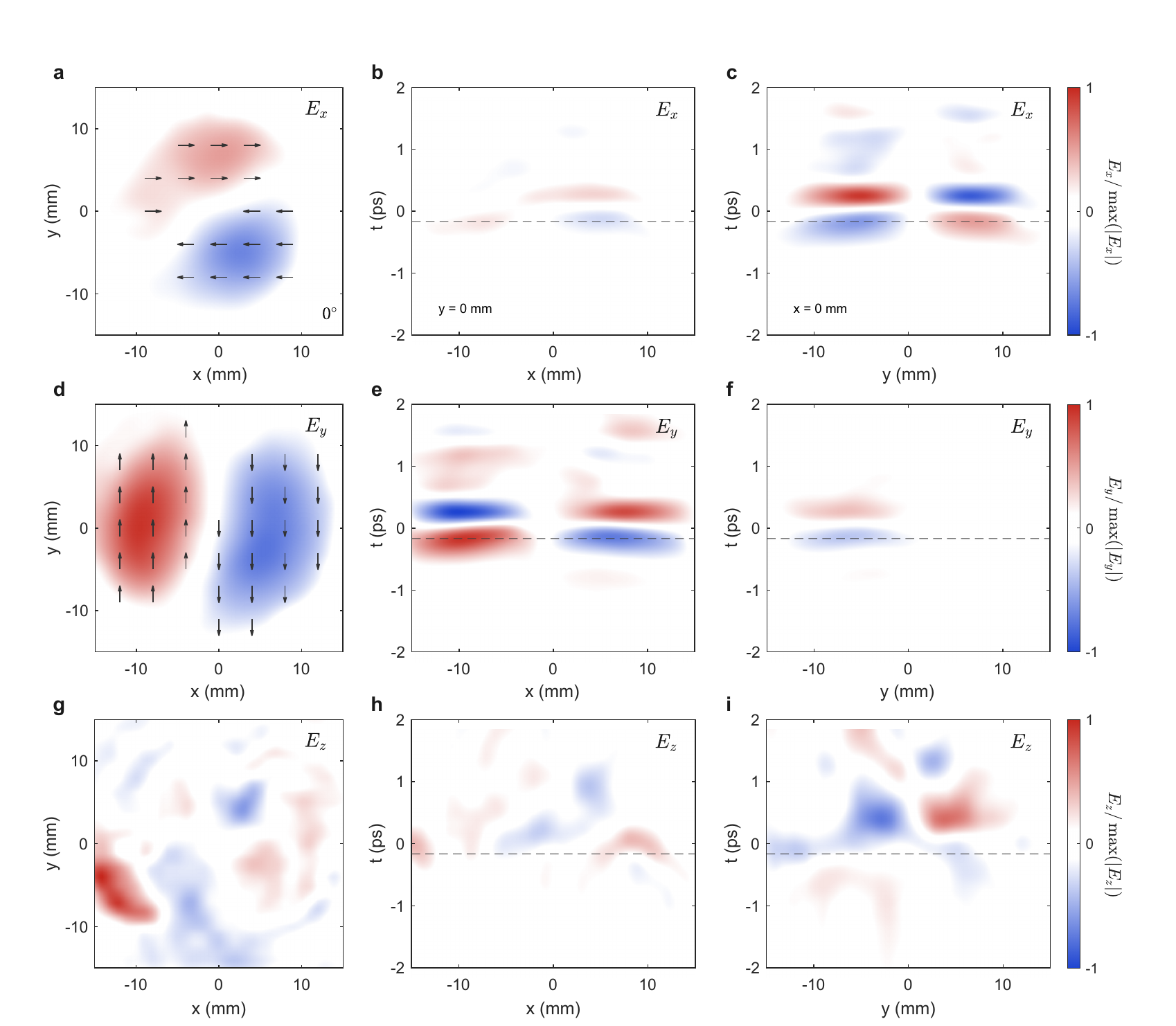}
    \caption{\textbf{Component-resolved spatiotemporal electric field for the $0^\circ$ pump-polarization state.}
    The rows show $E_x$, $E_y$, and the Maxwell-reconstructed $E_z$ component. The columns show the transverse $x$--$y$ distribution at a representative peak-field time and central $x$--$t$ and $y$--$t$ sections through the wavepacket. Each component is normalized independently. The orthogonally arranged bipolar patterns of $E_x$ and $E_y$ combine into the clockwise-azimuthal transverse field associated with the magnetic-dominant TE toroidal state. The temporal sections resolve the sign-changing single-cycle evolution of this azimuthal field; the corresponding toroidal vortex is carried by the reconstructed magnetic field $\mathbf{B}$ analyzed in Fig.\,3 of the main text.}
    \label{FigS_0deg_ExEyEz}
\end{figure*}

\begin{figure*}[t]
    \centering
    \includegraphics[width=0.95\textwidth]
    {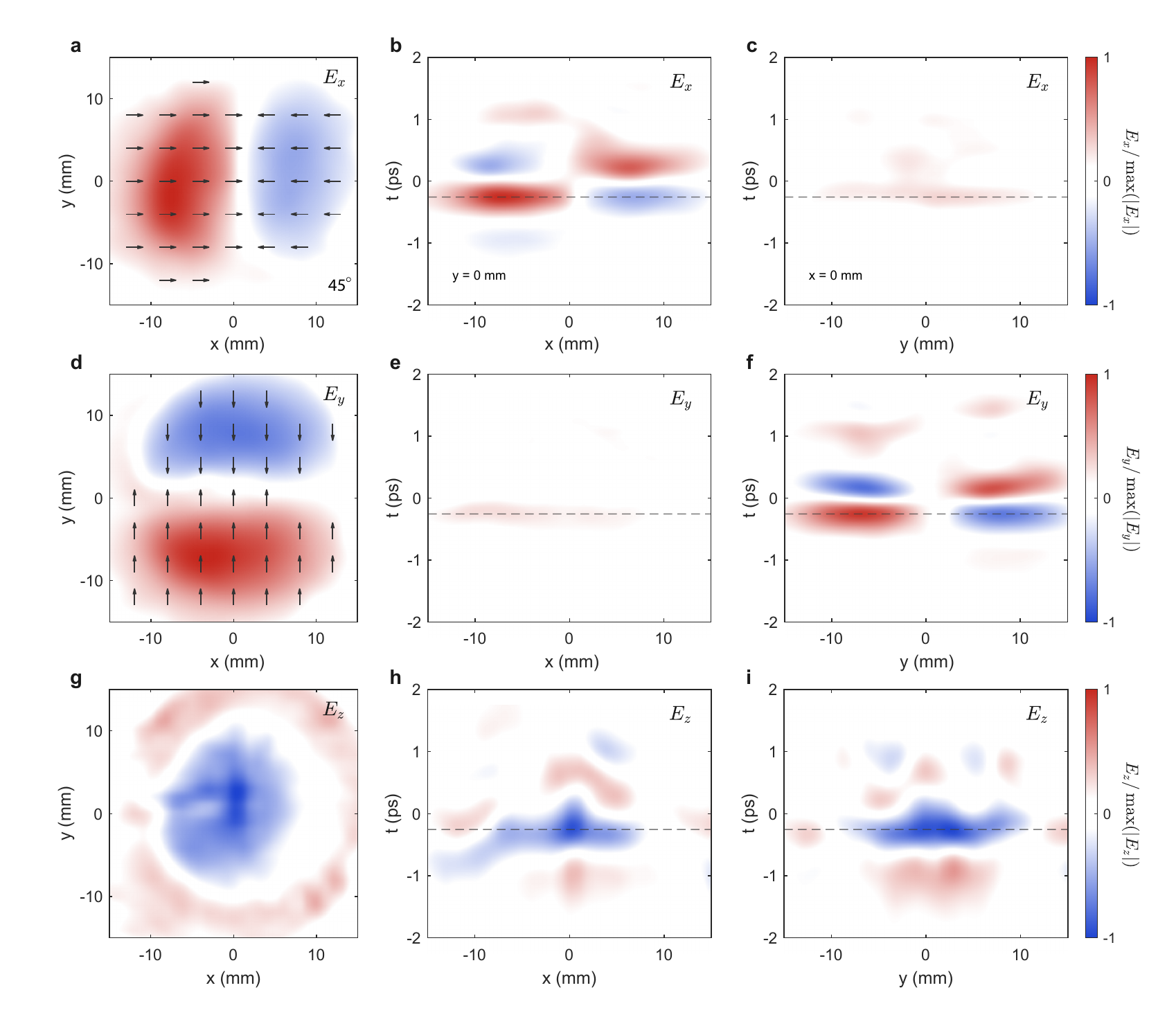}
    \caption{\textbf{Component-resolved spatiotemporal electric field for the $45^\circ$ pump-polarization state.}
    The rows show $E_x$, $E_y$, and the Maxwell-reconstructed $E_z$ component, and the columns show the transverse and two orthogonal space--time distributions. Each component is normalized independently. The Cartesian transverse components combine into the inward-radial field associated with the compressional photocurrent channel. The localized, sign-changing $E_z$ structure supplies the longitudinal sector of the meridional electric field. Together, the radial and longitudinal components constitute the electric-dominant TM toroidal state whose $\mathbf{E}$ vortex is traced and reconstructed in Fig.\,3 of the main text.}
    \label{FigS_45deg_ExEyEz}
\end{figure*}

\begin{figure*}[t]
    \centering
    \includegraphics[width=0.95\textwidth]
    {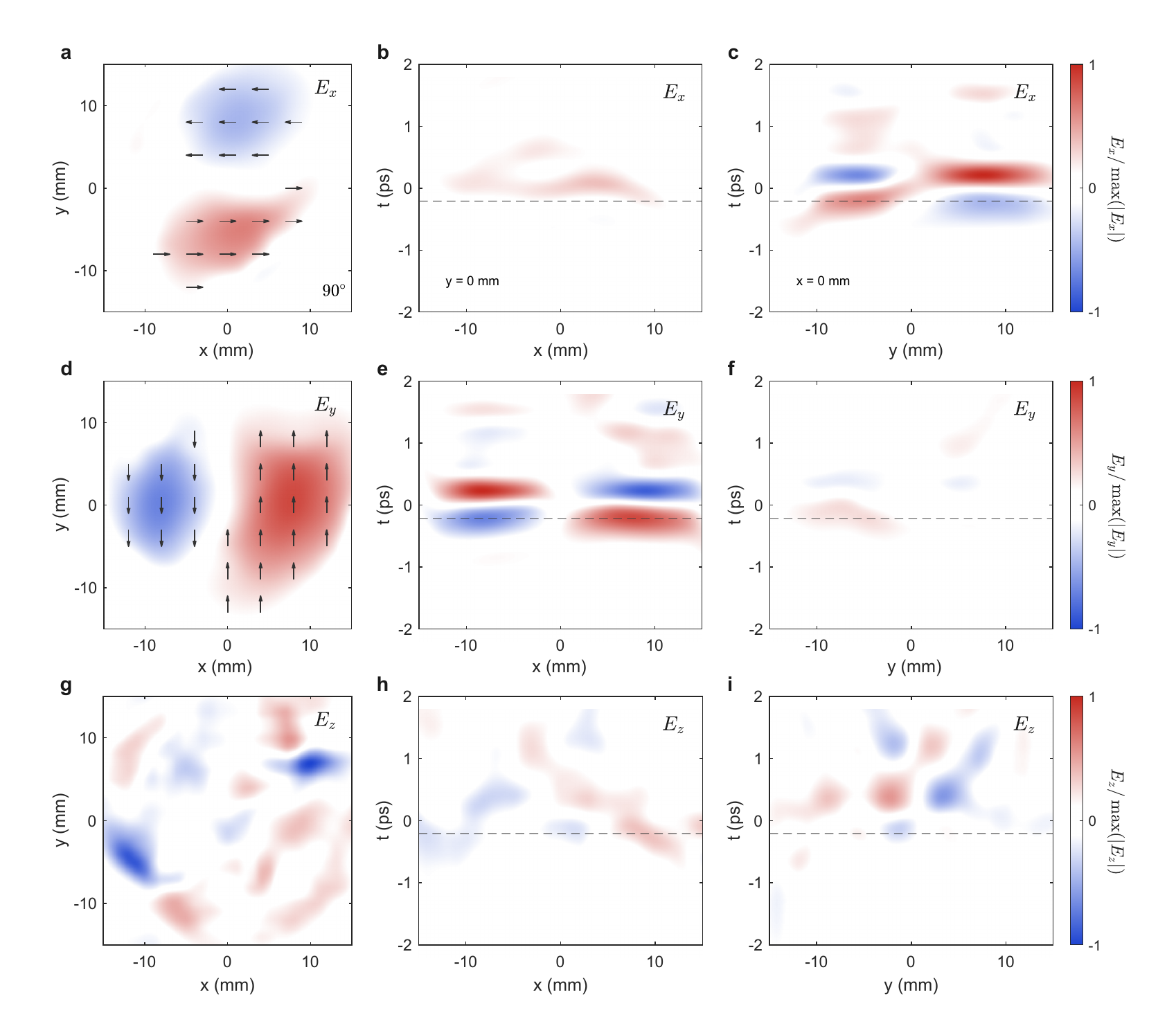}
    \caption{\textbf{Component-resolved spatiotemporal electric field for the $90^\circ$ pump-polarization state.}
    The rows and columns follow the arrangement in Fig.~\ref{FigS_0deg_ExEyEz}, with each component normalized independently. Relative to the $0^\circ$ excitation, the dominant positive and negative lobes of $E_x$ and $E_y$ are exchanged, producing the counter-clockwise-azimuthal transverse field. The corresponding reversal in the space--time sections is consistent with the $\pi$ phase shift between the two azimuthal states. This field is the negative-TE toroidal partner of the $0^\circ$ state, with its defining toroidal vortex resolved in the reconstructed magnetic field $\mathbf{B}$.}
    \label{FigS_90deg_ExEyEz}
\end{figure*}

\begin{figure*}[t]
    \centering
    \includegraphics[width=0.95\textwidth]
    {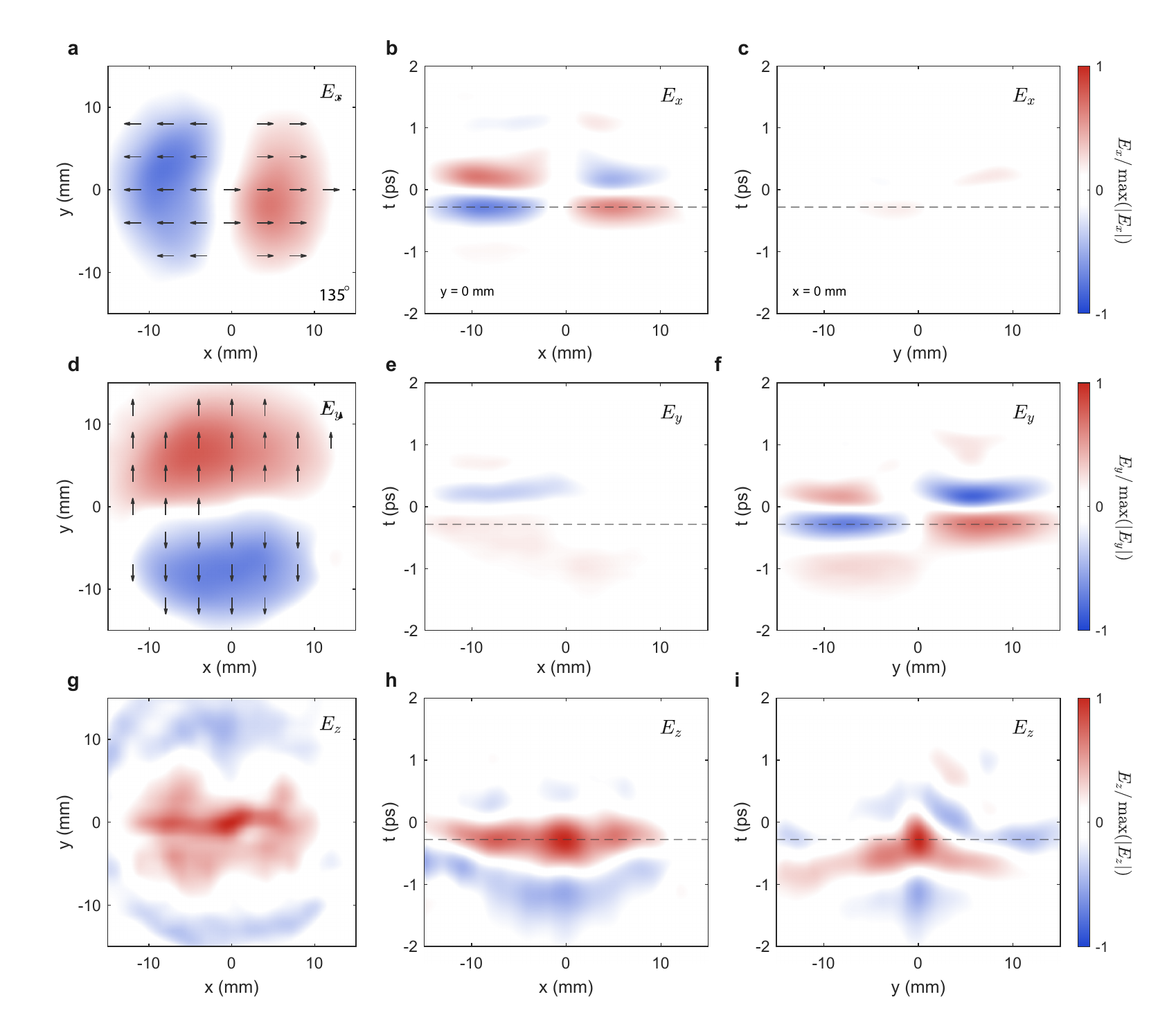}
    \caption{\textbf{Component-resolved spatiotemporal electric field for the $135^\circ$ pump-polarization state.}
    The rows show $E_x$, $E_y$, and the Maxwell-reconstructed $E_z$ component, while the columns show the transverse and two orthogonal space--time distributions. Each component is normalized independently. The transverse components form the outward-radial partner of the $45^\circ$ state, and the longitudinal component provides the complementary $E_z$ sector of the meridional field. Their combined space--time evolution is consistent with reversal of the toroidal-field orientation, identifying this excitation as the negative-TM partner of the $45^\circ$ state.}
    \label{FigS_135deg_ExEyEz}
\end{figure*}

To connect the polarization-controlled photocurrent channels to the toroidal and skyrmionic field structures discussed in the main text, Figs.\,\ref{FigS_0deg_ExEyEz}--\ref{FigS_135deg_ExEyEz} present the component-resolved electric fields of the four representative pump-polarization states. For each state, the rows show the measured transverse components, $E_x$ and $E_y$, and the Maxwell-reconstructed longitudinal component, $E_z$. The left column shows a transverse $x$--$y$ plane snapshot at a representative peak-field time, while the middle and right columns show the time-domain wavepacket as a function of $x$ and $y$ positions following the horizontal $y=0$ and the vertical $x=0$ slices, respectively. Note that these fields have been plotted with constant phase front of \cref{FigS_Spectra90deg}a, i.e., with a flat-field phase correction to offset the large curvature of the converging wavefronts. Each component is normalized to its own maximum absolute value. The color scales therefore represent field sign, spatial structure, and temporal phase evolution, rather than the relative amplitudes of different field components. In particular, the visible contrast of $E_z$, especially for the TE states, should not be interpreted as implying a longitudinal electric field comparable in magnitude to the transverse field.

The $0^\circ$ (\cref{FigS_0deg_ExEyEz}) and $90^\circ$ (\cref{FigS_90deg_ExEyEz}) excitations form the two azimuthal members of the tunable field family. In the transverse plane, the orthogonally arranged bipolar lobes of $E_x$ and $E_y$ combine to form clockwise- and counter-clockwise-azimuthal vector fields, respectively. The exchange of the positive and negative lobes between these two states reflects the $\pi$ phase reversal produced by rotating the pump polarization by $90^\circ$. The corresponding $x$--$t$ and $y$--$t$ sections show that this polarity reversal is maintained throughout the dominant single-cycle transient. These electric-field distributions are the transverse counterparts of the magnetic-dominant TE and negative-TE toroidal pulses. Accordingly, the defining meridional vortex of these two states is resolved in the reconstructed magnetic field, $\mathbf{B}$, in Fig.\,3 of the main text.

The $45^\circ$ (\cref{FigS_45deg_ExEyEz}) and $135^\circ$ (\cref{FigS_135deg_ExEyEz}) excitations form the complementary radial pair. Their leading pulse cycle $E_x$ and $E_y$ distributions combine into the inward- and outward-radial transverse states generated by the compressional photocurrent channels (which are opposite in polarity). In contrast to the azimuthal pair, the reconstructed $E_z$ component exhibits an organized longitudinal structure around the central transverse-field singularity, together with surrounding sign-changing features. The radial and longitudinal electric-field components jointly form the meridional field geometry characteristic of an electric-dominant TM toroidal pulse. Reversal of the radial current channel changes the orientation of this meridional field, producing the TM and negative-TM partners whose $\mathbf{E}$ vortex trajectories are analyzed in Fig.\,3 of the main text.

The $x$--$t$ and $y$--$t$ sections further show that the transverse lobes and longitudinal field evolve coherently across both space and time. Their coordinated sign reversals are confined to the principal THz transient, demonstrating that the measured structures are dynamical vector wavepackets rather than static transverse beam patterns. Together, the four datasets provide the component-level connection between the pump-selected photocurrent direction and the emitted toroidal mode. They therefore establish the experimental field basis for the TE/TM mode assignment, the three-dimensional vortex reconstruction in Fig.\,3 of the main text, and the subsequent analysis of ultrafast skyrmion number and helicity in Fig.\,4 of the main text. More broadly, the four states demonstrate two independent forms of optical control from a single metasurface: selection between magnetic and electric toroidal character and reversal of the corresponding toroidal polarity.

\begin{figure}[t]
    \centering
    \includegraphics[width=\linewidth]{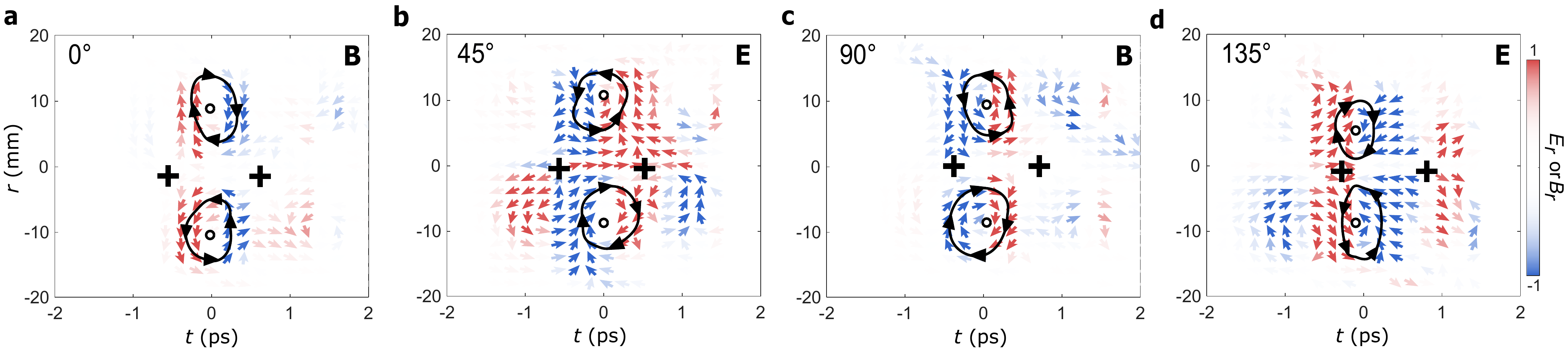}
    \caption{\textbf{Radial component representations of spatiotemporal fields.}
    Panels \textbf{a}--\textbf{d} correspond to pump-polarization angles of $0^\circ$,
    $45^\circ$, $90^\circ$, and $135^\circ$, respectively, and use the
    same coordinate sections as Fig.\,3b of the main text.}
    \label{FigS_Red-Blue}
\end{figure}

To provide a complementary view of the spatiotemporal vector fields in Fig.\,3 of the main text, in \cref{FigS_Red-Blue} we plot the four pump-polarization states using the \textit{r}-direction electric- or magnetic-field component, $E_r$ or $B_r$, via color mappings, whereas the field directions are indicated by arrows. This visualization directly reveals the temporal evolution of the radial field texture while retaining the local vector-field directions. This representation does not introduce an additional dataset or reconstruction.

The spatiotemporal skyrmionic textures of the four polarization states are characterized by the skyrmion number $N_{\mathrm{sk}}$ and helicity $\gamma$. The time-dependent skyrmion number is given by
\begin{equation}
    N_{\mathrm{sk}}(t)=\frac{1}{4\pi}
    \iint_{\Omega}
    \mathbf{n}\cdot
    \left(
        \frac{\partial\mathbf{n}}{\partial x}
        \times
        \frac{\partial\mathbf{n}}{\partial y}
    \right)
    \mathrm{d}x\,\mathrm{d}y,
    \label{eq:skyrmion_number}
\end{equation}
where $\mathbf{n}=\mathbf{F}/|\mathbf{F}|$ is the normalized instantaneous electric or magnetic field vector, and $\Omega$ denotes the transverse integration domain. For a skyrmionic texture with well-defined boundary conditions, the skyrmion number can be expressed as $N_{\mathrm{sk}}=pm$, where the polarity $p=\frac{1}{2}[n_z(0)-n_z(R)]$ describes the out-of-plane field reversal from the center to the boundary, and $m$ is the in-plane vorticity defined by
\begin{equation}
    m=\frac{1}{2\pi}\oint_{\mathcal{C}}
    \nabla_{\perp}\alpha\cdot\mathrm{d}\boldsymbol{l},
    \qquad
    \alpha=\operatorname{atan2}(F_y,F_x),
    \label{eq:vorticity}
\end{equation}
with $\mathcal{C}$ enclosing the texture center. Four states displayed in Fig.\,4 of the main text exhibit unit vorticity $m=+1$. Then, the opposite skyrmion-number signatures arise from polarity reversal.

The helicity $\gamma$ characterizes the in-plane field orientation relative to the radial direction and is defined as
\begin{equation}
    \gamma(\mathbf{r},t)=
    \alpha(\mathbf{r},t)-m\phi(\mathbf{r}),
    \qquad
    \phi=\operatorname{atan2}(y-y_c,x-x_c),
    \label{eq:helicity}
\end{equation}
where $(x_c,y_c)$ denotes the texture center. With $m=+1$, the polarization states with $\cos\gamma=+1$ and $-1$ exhibit radially outward and inward fields, respectively. Therefore, $N_{\mathrm{sk}}$ and $\gamma$ jointly characterize the polarity and in-plane orientation of the evolving skyrmionic field textures.
\clearpage

\clearpage

\clearpage

\clearpage

\clearpage

\clearpage

\section{S5. Vector vortex tracing for space-time toroidal pulse}\label{sec5}

The vortex loops in Fig.\,3 of the main text were extracted from the measured space-time field distributions by developing a vector vortex tracing algorithm. The same procedure was applied to all four pump-polarization states. For the TE toroidal states, the vortex loop was traced on the magnetic-field distributions, while the vortex loops of TM toroidal states were traced on the corresponding electric-field distributions.

The algorithm proceeds in two steps: elliptical optimization and Fourier normal deformation. The first step provides an optimized elliptical contour around the selected primary vortex point, while the second step locally deforms this contour to follow the measured vortex-vector direction. This two-stage procedure avoids direct contour extraction from noisy experimental data and constrains the final loop to remain smooth and localized.

In the elliptical optimization step, the space-time vector field map is first interpolated over the measurement plane. A tunable ellipse is then optimized around the selected vortex region, with its center, semi-axes, and in-plane rotation angle treated as fitting variables. For a trial loop with unit tangent $\boldsymbol{\tau}_i$, the angular mismatch to the local normalized vector direction $\mathbf{f}(\mathbf{r}_i)$ is defined as
\begin{equation}
    \alpha_i=\cos^{-1}
    \left(
    \left|
    \boldsymbol{\tau}_i\cdot\mathbf{f}(\mathbf{r}_i)
    \right|
    \right).
\end{equation}
The absolute value accounts for the head-to-tail ambiguity of the local vortex-vector direction, considering both parallel and antiparallel alignments as tangential. The cost function aims to minimize the mean-squared angular mismatch between the ellipse tangent and the local vector direction, with a weak contribution from the maximum mismatch. Furthermore, a mild regularization is applied to avoid unnecessary expansion and physically invalid aspect ratios. The optimization can be repeated for several initial ellipse orientations, and the lowest-cost solution is retained as the reference ellipse contour.

In the Fourier-deformation step, the optimized ellipse is deformed only along its local normal direction. These normal displacements are described by a bounded Fourier series, which limits the maximum displacement and also prevents the trace from moving far outside the local vortex region. The cost function for this step also minimizes the mean and maximum angular mismatches to the interpolated vector field. Moreover, it imposes additional penalties on the displacement amplitude and high-order Fourier coefficients. The high-order penalty suppresses small oscillations while maintaining the main deformation. During the deformation, multiple Fourier-coefficient seeds are calculated, and the lowest-cost deformed curve is selected as the final trace.

\begin{figure}[t]
    \centering
    \includegraphics[width=0.75\linewidth]{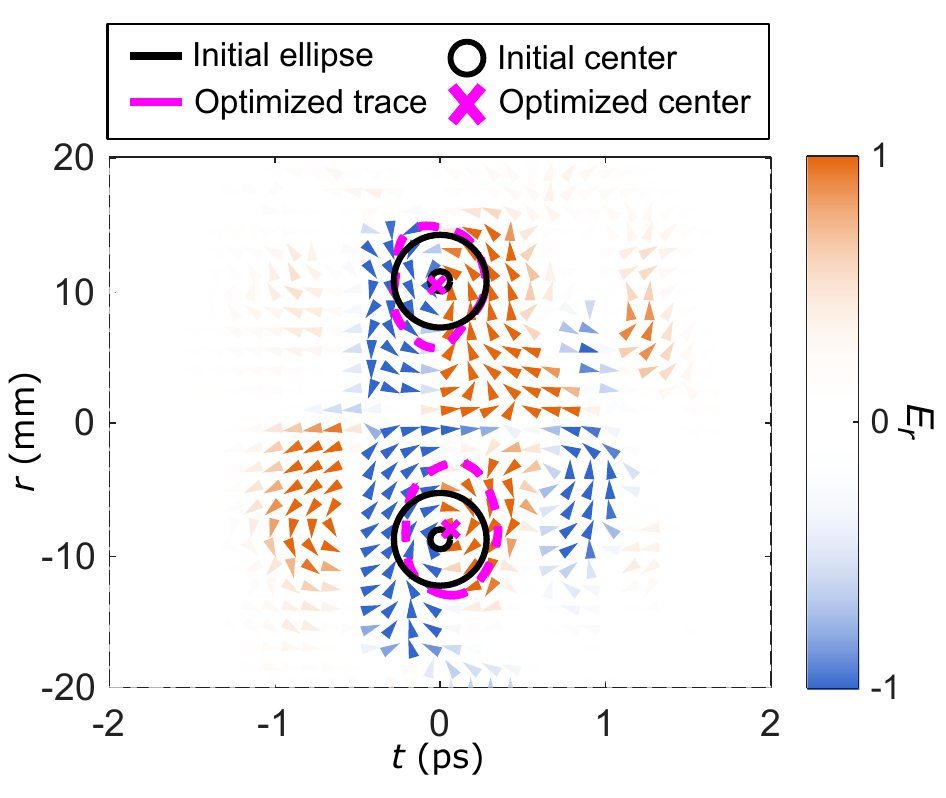}
    \caption{\textbf{Vector vortex tracing.}
    The vortex-field tracing method is illustrated by the electric field under 45$^\circ$ incident polarization. The black solid line and open circle denote the initial elliptical trajectory and its center, respectively. The magenta dashed line and cross denote the optimized trajectory and its center, respectively.}
    \label{FigS_TracingMethod}
\end{figure}

\cref{FigS_TracingMethod} illustrates the vortex tracing method for the electric-field map measured under $45^\circ$ incident polarization. The optimized traces and centers are presented by magenta dashed curves and crosses, respectively, benchmarking the initial elliptical contours and centers in black solid curves and open circles. The optimized loop exhibits a closer fit to the local vortex-vector direction than the initial ellipses.

The mathematical basis of the proposed tracing method can be further described by parameterizing the final vortex loop as
\begin{equation}
    \mathbf{r}(\theta)=\mathbf{r}_0(\theta)+\delta(\theta)\mathbf{n}_0(\theta),
\end{equation}
where $\mathbf{r}_0(\theta)$ is the optimized reference ellipse, $\mathbf{n}_0(\theta)$ is its unit normal vector, and $\delta(\theta)$ is the Fourier normal displacement. Besides the local perturbation $\delta(\theta)\mathbf{n}_0(\theta)$, the optimized ellipse $\mathbf{r}_0(\theta)$ also determines the final loop $\mathbf{r}(\theta)$ and the reconstructed three-dimensional toroids. This reference ellipse is partially constrained by the initial ellipse, 
\begin{equation}
    \mathbf{r}_{\mathrm{ini}}(\theta)=\mathbf{c}_{\mathrm{ini}}+
    R(\psi_{\mathrm{ini}})
    \begin{bmatrix}
        T_{\min}\cos\theta\\
        R_{\min}\sin\theta
    \end{bmatrix},
\end{equation}
where $\mathbf{c}_{\mathrm{ini}}$, $\psi_{\mathrm{ini}}$, $T_{\min}$, and $R_{\min}$ are the initial center, in-plane rotation angle, temporal semi-axis, and spatial semi-axis, respectively.

\begin{figure*}[!t]
    \centering
    \includegraphics[width=\linewidth]{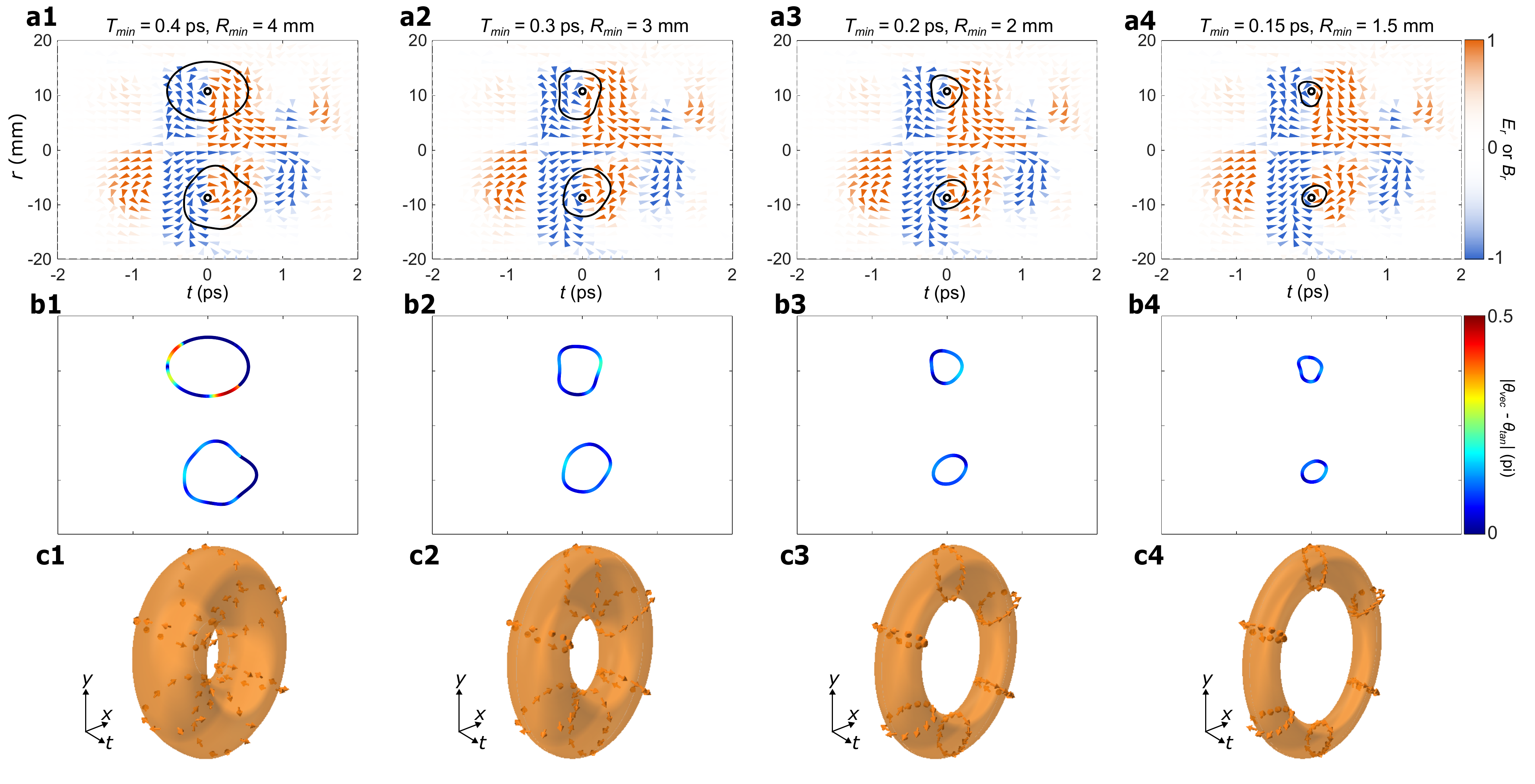}
    \caption{\textbf{Three-dimensional reconstruction across varying initial geometries.}
    \textbf{a}, Trade-off between tracing accuracy and reconstruction volume $(T_{\min},R_{\min})$. The black curves denote the optimized vortex loops overlaid on the measured space--time electric-field maps.
    \textbf{b}, Local angular mismatch, $|\theta_{\mathrm{vec}}-\theta_{\mathrm{tan}}|$, along the corresponding optimized loops.
    \textbf{c}, Three-dimensional toroids reconstructed from the traced vortex loops. Larger initial semi-axis pairs produce larger reconstructed volumes but also increase the tracing error, revealing a trade-off between reconstruction volume and tracing accuracy.}
    \label{FigS_TracingSettingSweep}
\end{figure*}

The remaining tracing settings and the subsequent Fourier normal deformation procedure were kept unchanged for four different initial geometries. As shown in \cref{FigS_TracingSettingSweep}, larger initial semi-axis pairs yield larger reconstructed toroidal structures but also increase the local tangential angle errors along the traced loops. In contrast, smaller initial ellipses yield lower tracing errors but reconstruct more compact toroidal structures. This comparison reveals a trade-off between reconstruction volume and tracing accuracy. Therefore, the traced vortex loops in Fig.~3 of the main text were selected as the largest reconstructed loops with tracing errors below $0.125\pi$ under varying initial ellipse geometries.
\clearpage

\clearpage

\clearpage

\section{S6. Space-time non-separability}\label{sec6}

\begin{figure*}
    \centering
    \includegraphics[width=0.95\textwidth]{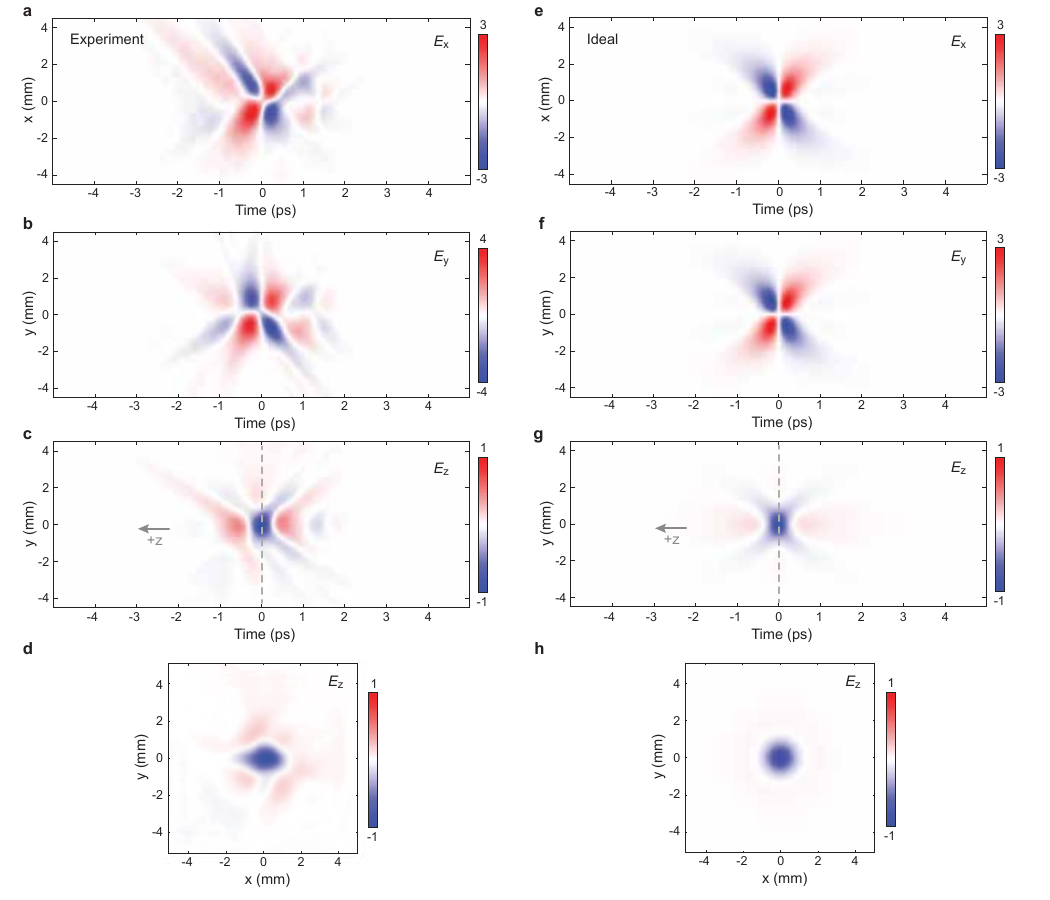}
    \caption{\textbf{Focal plane distribution following numerical propagation.}
    \textbf{a}, Measured $E_x$ and \textbf{b}, $E_y$ electric field distributions following numerical propagation to the focal plane for representative 45$^\circ$ dataset. \textbf{c}, Corresponding $E_z$ distribution at the focal plane. \textbf{d}, Spatial $xy$ cross-section of the $E_z$ field at the focal plane. \textbf{e--h}, Ideal toroidal pulse cross-sections for comparison.}
    \label{FigS_Propagated45deg}
\end{figure*}

To verify the pulse characteristics at different propagation planes, we utilize the complete characterization performed approximately 7\,cm away from the focal plane to numerically propagate the field (see \cref{FigS_ExperimentSchematic}). After spatial and temporal Fourier transforms of the $E_x$ and $E_y$ field components, we propagate forward in $z$ via $e^{-i k_z z}$ phase factors before performing inverse transforms. Propagated pulse cross-sections are shown in \cref{FigS_Propagated45deg}a,b for a representative compressional mode driven at 45$^\circ$ incident polarization, shown at the focal plane. These reveal a nearly single-cycle pulse, with an inward-radial leading peak cycle consistent with the outward-flowing current illustrated in Fig.\,2 of the main text, where $\mathbf{E}_\mathrm{THz} \propto -d\mathbf{j}/dt$. The pulse shape is also in close agreement with the ideal calculated values in \cref{FigS_Propagated45deg}e,f. Utilizing Gauss’s law to determine $E_z$, we find the expected central peak, pointing against the propagation direction and aligned with the central node in $E_x$ and $E_y$. These further clarify the toroidal pulse characteristics shown in Fig.\,3 of the main text, now at the focal plane.
    
\begin{figure*}
    \centering
    \includegraphics[width=0.9\textwidth]{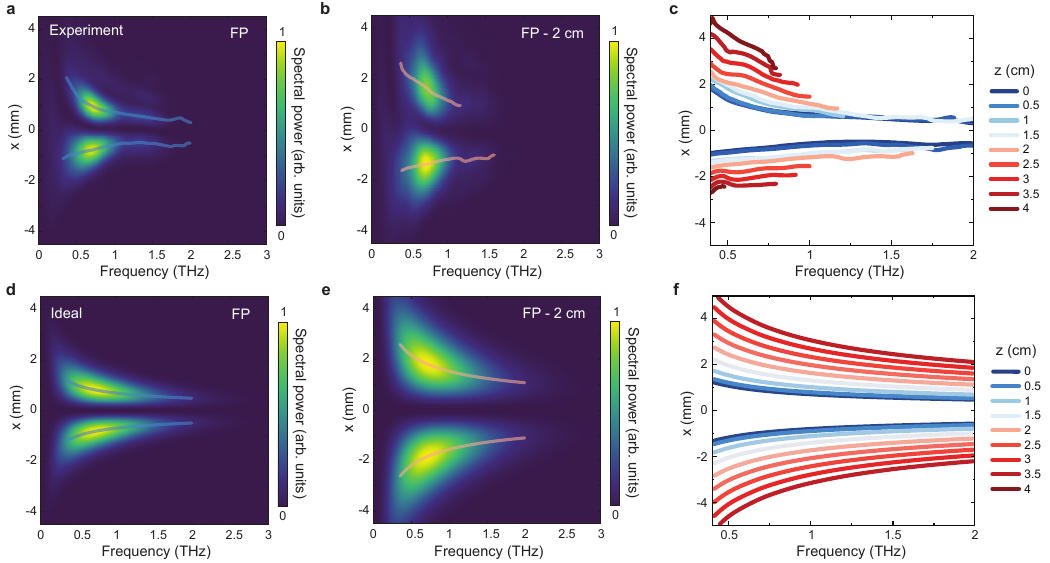}
    \caption{\textbf{Space-time non-separability.}
    \textbf{a}, Spatially-resolved fast-Fourier transform (FFT) power spectrum, displayed at the focal plane (FP) for representative 135$^\circ$ dataset. Solid curves show fit spatial peaks at each frequency. \textbf{b}, Spatially-resolved FFT shown 2\,cm before the FP. \textbf{c}, Summary of frequency-dependent spatial peaks at a series of propagation distances away from (before) the focal plane. \textbf{d--f}, Corresponding analytic curves for an ideal toroidal pulse.}
    \label{FigS_Isodiffraction}
\end{figure*}

Ideal toroidal pulses are space-time non-separable solutions to Maxwell’s equations\cite{Hellwarth1996} where higher frequencies are concentrated at smaller radii. While this is a characteristic feature of diffraction-limited focusing around the focal plane, toroidal pulses preserve the spatially-dependent frequency profile during propagation, with the same relative divergence for each frequency component\cite{Zdagkas2020}. This isodiffraction phenomenon is tested through the verified numerical propagation procedure. At the focal plane, a temporal fast-Fourier transform of the 135$^\circ$ $E_x$ distribution shows the spatially-dependent frequency content (\cref{FigS_Isodiffraction}a) expected for an ideal toroidal mode (\cref{FigS_Isodiffraction}d). This behavior is preserved but broadens out during propagation, as shown 2\,cm away from the focal plane (\cref{FigS_Isodiffraction}b,e). Frequency-dependent fits to the spatial peaks clarify this behavior, summarized for a series of propagation planes (\cref{FigS_Isodiffraction}c,f; datasets are truncated due to progressively lower signal-to-noise ratio at high frequencies). This verifies the space-time non-separability and shows trends consistent with isodiffraction.
\clearpage

\clearpage

\clearpage

\bibliography{References}